\documentclass[
  aps, 
  prb, 
  showkeys,
  nobibnotes,
  reprint,
]
  {revtex4-2}

\usepackage{amssymb}
\usepackage{amsmath}
\usepackage{mathtools}
\usepackage{stmaryrd}

\usepackage{mathrsfs}

\usepackage{xfrac}

\usepackage{adjustbox}
\usepackage{tcolorbox}
\newtcolorbox{standout}{
  colback=gray!15,
  boxrule=0pt,
  left=.3cm,
  right=.3cm,
  top=.18cm,
  bottom=.18cm,
  boxsep=0pt
}

\usepackage{amsthm}
\newtheorem{theorem}{Theorem}[subsection]

\newtheorem{proposition}[theorem]{Proposition}
\theoremstyle{definition}
\newtheorem{definition}[theorem]{Definition}
\newtheorem{example}[theorem]{Example}

\usepackage{tikz}
\usetikzlibrary{
  cd,
  arrows.meta,
  backgrounds,
  decorations.pathmorphing,
  calc
}

\tikzcdset{
  arrow style=tikz, 
  diagrams={
    >={
      Computer Modern Rightarrow[
        length=4pt, width=4pt
      ]
    },
    shorten=0pt
  },
  hook/.style={
    {Hooks[right, length=2pt, width=8pt]}->
  }
}

\definecolor{darkblue}{rgb}{0.05,0.25,0.65}
\definecolor{darkgreen}{RGB}{20,140,10}
\definecolor{lightgray}{rgb}{0.9,0.9,0.9}
\definecolor{darkorange}{RGB}{200,100,5}
\definecolor{darkyellow}{rgb}{.91,.91,0}
\definecolor{lightolive}{RGB}{225, 220, 185}

\usepackage[
  colorlinks=true,
  linkcolor=darkgreen,
  citecolor=darkgreen,
  urlcolor=darkgreen
]{hyperref}

\makeatletter
\newcommand{\acomment}{%
  \@ifnextchar\bgroup{\@processcomment}{}%
}
\def\@processcomment#1{%
  \scalebox{.7}{\color{black!95}#1}%
  \@ifnextchar\bgroup{%
    \\\@processcomment
  }{}%
}
\makeatother

\tikzset{
  snake left/.style={
    rounded corners,
    to path={
      let \p1 = (\tikztostart.east),
          \p2 = (\tikztotarget.west),
          \p3 = ($(\p1)!0.5!(\p2)$),
          \n1 = {8pt} 
      in
      (\p1)
      -- (\x1 + \n1, \y1)
      -- (\x1 + \n1, \y3)
      -- (\x2 - \n1, \y3) \tikztonodes
      -- (\x2 - \n1, \y2)
      -- (\p2)
    }
  }
}

\tikzset{
  uphordown/.style={
    rounded corners,
    to path={
      let \p1 = (\tikztostart.north),
          \p2 = (\tikztotarget.north),
          \n1 = {max(\y1,\y2) + 8pt}
      in
      (\p1)
      -- (\x1, \n1)
      -- (\x2, \n1) \tikztonodes 
      -- (\p2)
    }
  }
}

\tikzset{
  downhorup/.style={
    rounded corners,
    to path={
      let \p1 = (\tikztostart.south),
          \p2 = (\tikztotarget.south),
          \n1 = {min(\y1,\y2) - 8pt}
      in
      (\p1)
      -- (\x1, \n1)
      -- (\x2, \n1) \tikztonodes 
      -- (\p2)
    }
  }
}

\tikzset{
  rightvertleft/.style={
    rounded corners,
    to path={
      let \p1 = (\tikztostart.east),
          \p2 = (\tikztotarget.east),
          \n1 = {max(\x1,\x2) + 8pt}
      in
      (\p1)
      -- (\n1, \y1)
      -- (\n1, \y2) \tikztonodes 
      -- (\p2)
    }
  }
}

\tikzset{
  leftvertright/.style={
    rounded corners,
    to path={
      let \p1 = (\tikztostart.west),
          \p2 = (\tikztotarget.west),
          \n1 = {min(\x1,\x2) - 8pt}
      in
      (\p1)
      -- (\n1, \y1)
      -- (\n1, \y2) \tikztonodes 
      -- (\p2)
    }
  }
}

\newcommand{\grayoverbrace}[2]{%
  \mathcolor{gray}{\overbrace{%
    \mathcolor{black}{#1}%
  }^{\mathcolor{black}{#2}}}%
}

\newcommand{\grayunderbrace}[2]{%
  \mathcolor{gray}{\underbrace{%
    \mathcolor{black}{#1}%
  }_{\mathcolor{black}{#2}}}%
}

\newcommand{\defneq}{\equiv}

\newcommand{\weakHomotopyEquivalence}{\sim}

\newcommand{\Integers}{\mathbb{Z}}

\newcommand{\CyclicGroup}[1]{
  \Integers_{\mskip-3mu/\mskip-1mu{#1}\mskip-.5mu}
}

\newcommand{\ZThree}{%
  \CyclicGroup{3}%
}

\newcommand{\SpaceGroup}[1]{\mbox{#1}}

\newcommand{\acts}{\hspace{-.5pt}\raisebox{1.4pt}{\;\rotatebox[origin=c]{90}{$\curvearrowright$}}\hspace{1pt}}

\newcommand{\GeneralBrillouinTorus}[1]{\widehat{T}^{#1}}

\newcommand{\BrillouinTorus}{\GeneralBrillouinTorus{2}}

\newcommand{\ValenceBundle}{\mathcal{V}}

\newcommand{\HilbertSpace}{\mathcal{H}}

\newcommand{\ClassifyingMap}{H/\vert H \vert}

\newcommand{\CrystalMomentum}
{[\vec{\rule{0pt}{6pt}\smash{k}}\hspace{1pt}]\hspace{-0pt}}

\newcommand{\TriVertex}[2]{
\begin{scope}[shift={(#1:#2)}]
\draw[
  line width=1.4,
  color=lightgray
]
  (0:0) --
  (0:1.5);
\draw[
   -Latex,
   gray!70,
   line width=.6
]
  (0:.9) --
  (0:.9+.01);
\node[scale=.3] 
  at (0:.62)  
  {{\hspace{-2.2pt}\color{gray}$v$\hspace{-2.2pt}}};

\draw[
  line width=1.4,
  color=lightgray
]
  (120:0) --
  (120:1.5);
\draw[
   -Latex,
   gray!70,
   line width=.7
]
  (120:.9) --
  (120:.9+.01);
\node[scale=.3] 
  at (120:.62)  
  {\rotatebox[origin=c]{-60}{{\hspace{-2.2pt}\color{gray}$w$\hspace{-2.2pt}}}};

\draw[
  line width=1.4,
  color=lightgray
]
  (240:0) --
  (240:1.5);
\draw[
   -Latex,
   gray!70,
   line width=.7
]
  (240:.9) --
  (240:.9+.01);
\node[scale=.33] 
  at (240-.8:.42)  
  {\rotatebox[origin=c]{60}{{\hspace{-2.2pt}\color{gray}$-w-v$\hspace{-2.2pt}}}};
  
\fill[
  lightgray
]
  (0,0) circle (.08);
\fill[
  lightgray
]
  (0:1.5) circle (.08);
\fill[
  lightgray
]
  (120:1.5) circle (.08);
\fill[
  lightgray
]
  (240:1.5) circle (.08);

\end{scope}

}

\newcommand{\HeisenbergGroup}[1]{%
 \mathbb{Z}^2 \times_2 \CyclicGroup{#1}%
}

\newcommand{\hotype}[1]{\mathcal{#1}}

\newcommand{\inlinetikzcd}[1]{\begin{tikzcd}[sep=small, ampersand replacement=\&]#1\end{tikzcd}}

\def\arraystretch{1.3}

\begin{document}

\setlength{\abovedisplayskip}{3.5pt}
\setlength{\belowdisplayskip}{3.5pt}
\setlength{\abovedisplayshortskip}{-5pt}
\setlength{\belowdisplayshortskip}{2pt}



\author{Hisham Sati}
\email{hsati@nyu.edu}
\affiliation{Center for Quantum and Topological Systems, Division of Science, New York University, Abu Dhabi, UAE}


\author{Urs Schreiber}
\email{us13@nyu.edu}
\affiliation{Center for Quantum and Topological Systems, Division of Science, New York University, Abu Dhabi, UAE}

\title{
  Fragile  Topological Phases 
    and 
  Topological Order 
    of  
  2D Crystalline Chern Insulators
}

\date{\today}

\begin{abstract}
  We apply methods of equivariant homotopy theory, which may not previously have found due attention in condensed matter physics, to classify first the fragile/unstable topological phases of 2D crystalline Chern insulator materials, and second the possible topological order of their fractional cousins. We highlight that the phases are given by the \emph{equivariant 2-Cohomotopy} of the Brillouin torus of crystal momenta (with respect to wallpaper point group actions) ---  which, despite the attention devoted to crystalline Chern insulators, seems not to have been considered before. Arguing then that any topological order must be reflected in the \emph{adiabatic monodromy} of gapped quantum ground states over the covariantized space of these band topologies, we compute the latter in examples where this group is non-abelian, showing that any potential \emph{FQAH anyons} must be localized in momentum space. We close with an outlook on the relevance for the search for topological quantum computing hardware. Mathematical details are spelled out in a supplement \cite{supp}.
\end{abstract}


\maketitle


\tableofcontents

\section{Agenda}
\label{Agenda}

Since we introduce novel perspectives, we start here with a brief outline before discussing more details in \S\ref{MotivationAndIntroduction}.

\subsection{Fragile Crystalline Chern Phases}

Crystalline Chern insulators 
(CI, cf. \cite{AndoFu2015}\cite[\S8-9]{Sergeev2023})
are the much-studied archetype of topological quantum materials \cite{GuptaSaxena18, Wang2020}, 
and yet it appears that the \emph{fragile} topological phases
(in the terminology of \cite{PoHarukiVishwanath2018,PoZouSenthilVishwanath2019,Bouhon2020,BrouwerDwivedi2023})
of the most basic case of crystalline \emph{2-dimensional 2-band} systems --- of which the celebrated \emph{Haldane model} \cite[\S 8.3]{Sergeev2023}  is a prime example --- have not been really classified before:

First, disregarding crystal symmetries, it is well-known (cf. \cite[(8.3-4)]{Sergeev2023} and \S\ref{MotivationAndIntroduction} below) that such topological phases are labeled by homotopy classes of maps 
\begin{equation}
  \label{MapInIntroduction}
  \begin{tikzcd}
    \BrillouinTorus
    \ar[rr, "{ H/\vert H \vert }"]
    &&
    S^2
  \end{tikzcd}
\end{equation}
from the Brillouin torus $\BrillouinTorus$ of crystal momenta (cf. \cite[\S 2.1]{Thiang2025}) to the 2-sphere $S^2$ of normalized relative $2 \times 2$ Bloch Hamiltonians $H$.

We highlight
(with \cite{SS25-FQAH,SS25-FQH}, surveyed in \cite{SS25-ISQS29}),
that these classes constitute the \emph{2-Cohomotopy}
\cite[\S VII]{STHu59}\cite[Ex. 2.7]{FSS23-Char} of the Brillouin torus, an unstable or \emph{nonabelian} \cite[\S 2]{FSS23-Char} cohomology refinement of the more popular but also coarser (stable) K-theory classification (for which cf. \cite{Stehouwer2021}\cite[\S 2]{SS22-Ord}).

However, it is only in this simple situation \eqref{MapInIntroduction} that 2-Cohomotopy coincides with ordinary integral cohomology as well as with K-theory, all of which give here that the topological phase is indexed by the \emph{winding number} of \eqref{MapInIntroduction} or equivalently by the (first and only) eponymous \emph{Chern number} $C \in \mathbb{Z}$ of the crystal's valence line bundle \eqref{TheUniversalLineBundle}. But this coincidence of cohomology classifications is crucially lifted when more details are brought into the picture:

Namely it is clear --- though rarely stated; we pointed this out in \cite{SS25-FQAH} and will again explain it in \S\ref{MotivationAndIntroduction} below --- that \emph{crystalline $G$-symmetry protection} 
\cite[(9.1)]{Sergeev2023}
constrains the normalized Bloch Hamiltonians \eqref{MapInIntroduction} 
to be maps \emph{equivariant} (cf. \cite[\S1.2]{Bredon1972})
with respect to suitable $G$-actions on both sides:
\begin{equation}
  \label{EquivariantMapInIntroduction}
  \begin{tikzcd}
    \BrillouinTorus
    \ar[
      in=34+90, out=-34+90,
      looseness=3.5,
      "{ G }"{description}
    ]
    \ar[rr, "{ H/\vert H \vert }"]
    &&
    S^2
    \mathrlap{\,,}
    \ar[
      in=34+90, out=-34+90,
      looseness=4.4,
      "{ G }"{description}
    ]
  \end{tikzcd}
\end{equation}
whence the topological phases are now \emph{equivariant homotopy classes} (cf. \cite[p. 5]{tomDieck1987}) constituting the \emph{equivariant 2-Cohomotopy} \cite[Ex. 4.5.8]{SS25-Bun}\cite[\S 6]{SS26-Orb}\cite[(3)]{SS20-Tad}\cite{BSS21-Frac} of the Brillouin torus. 

Obvious as this may be, these fragile crystalline topological phases seem not to have been computed or even considered before. Mathematically, they are the connected components $\pi_0$ of the $G$-equivariant $(-)^G$ mapping space $\mathrm{Map}(-,-)$, in that:
\begin{equation}
  \label{FragilePhasesIdeaInIntro}
  \left\{
    \adjustbox{
      scale=0.8,
      margin=2pt,
      bgcolor=lightgray!70
    }{%
    \def\arraystretch{1}%
    \begin{tabular}{@{}c@{}}%
      \bf 
      Fragile topological phases
      \\
      of $G$-symmetry protected
      \\
      2D 2-band Chern insulators
    \end{tabular}%
    }%
  \right\}
  =
  \pi_0 
  \,
  \mathrm{Map}\big(
    \BrillouinTorus
    ,
    S^2
  \big)^{\!G}
  .
\end{equation}
We compute this here, as $G$ ranges over point groups of 2D crystallographic groups. The nature of the result seems not to have been anticipated before: For symmorphic space groups and rotational band symmetries we find \cite[\S IV.B]{supp} that the topological phases \eqref{FragilePhasesIdeaInIntro} are generically labeled by \emph{pairs} $(C,c)$ consisting (i) of an integer Chern number $C$, divisible by the order $o$ of the point group, and
(ii) of an element $c$ of a finite set $[n]$ of (diffeomorphism-invariant) ways that the high-symmetry points of the Brillouin torus map to the fixed poles in $S^2$ (cf. Fig. \ref{TheFixedPoles}):
\begin{equation}
  \label{GenericFormOfFragilePhases}
  (C,c)
  \in
  o\mathbb{N} \times [n]
  \,.
\end{equation}

The divisibility of the Chern number by some order $o$ may have been anticipated before, but the second factor $[n]$ means that there may be more distinct phases than previously recognized.

\subsection{Topological Order of FCIs}
\label{OnTopologicalOrderOfFCIs}

But we go further:
Rather more recently, \emph{fractional} Chern insulators (FCI) exhibiting a
fractional quantum anomalous Hall effect (FQAH, theoretically predicted and developed in \cite{neupert2011fractional,parameswaran2013fractional,roy2014band,regnault2011fractional}) have been observed in laboratories \cite{Cai2023, Zeng2023, Park2023, Lu2024, Zhang2025}. Very recently, authors have been looking to understand
their expected anyonic topological order \cite{KobayashiEtAl2025,Shi2025}  (the former in view of crystalline symmetry protection, cf.  \cite{LuEtAl2012}) --- but the fundamental question remains: 

\begin{standout}
What is the nature and phenomenology of anyonic topological quantum states in crystalline 2D FCIs exhibiting an FQAH effect? How do they manifest, and how are they classified?
\end{standout}

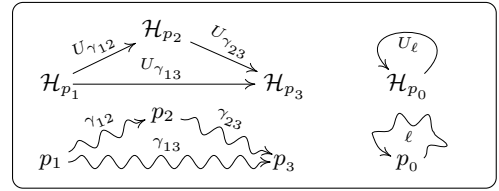
\begin{figure}[htb]
\centering
\caption{
  \label{AdiabaticQuantumTransport}
  Under adiabatic tuning of classical parameters $p$ along paths $\gamma$ in parameter space, the ground states of a gapped quantum system $\HilbertSpace$ undergo unitary transformations $U_\gamma$. For topological states, these $U_\gamma$ depend only on the homotopy class of $\gamma$ (relative endpoints), thus making the Hilbert spaces $\HilbertSpace_p$ constitute a \emph{local system} or \emph{flat bundle} over parameter space, hence equivalently a linear representation of the \emph{fundamental group} of closed paths (loops) $\ell$ at any base point $p_0$:
}
\centering
\adjustbox{
  rndfbox=4pt
}{
$
\begin{tikzcd}[
  decoration=snake,
]
  &
  \HilbertSpace_{p_2}
  \ar[
    dr,
    shorten=-2pt,
    "{ U_{\gamma_{{}_{23}}} }"{sloped}
  ]
  \\[-11pt]
  \HilbertSpace_{\mathrlap{p_{{}_{1}}}}
  \ar[
    rr,
    shorten >=-1pt,
    black,
    "{ U_{\gamma_{{}_{13}}} }"
  ]
  \ar[
    ur,
    shorten >=-2pt,
    "{
      U_{\gamma_{{}_{12}}}
    }"{sloped}
  ]
  &&
  \HilbertSpace_{p_{{}_3}}
  &
  \HilbertSpace_{p_{{}_0}}
  \ar[
    in=45+90,
    out=-45+90,
    looseness=4.7,
    shift left=2pt,
    shorten <=-2pt,
    "{ U_\ell }"
  ]
  \\[-20pt]
  &
  p_2
  \ar[
    dr,
    decorate,
    shorten <=-2pt,
    shorten >=-4pt,
    "{ \gamma_{{}_{23}} }"{yshift=2pt, sloped}
  ]
  &
  \\[-11pt]
  p_1
  \ar[
    rr, 
    decorate,
    shift right=1pt,
    shorten <=-2pt,
    shorten >=-3pt,
    "{
      \gamma_{{}_{13}}
    }"{yshift=2pt}
  ]
  \ar[
    ur,
    decorate,
    shorten <=-2pt,
    shorten >=-2pt,
    "{ 
      \gamma_{{}_{12}} 
    }"{yshift=2pt, sloped}
  ]
  &&
  p_{{}_3}
  &
  p_{{}_0}
  \ar[
      in=52+90,
      out=-52+90,
      looseness=5,
      shift left=4pt,
      decorate,
      shorten <=-1pt,
      "{ \ell }"
  ]
\end{tikzcd}
$
}
\end{figure}

To understand what topological quantum theory predicts in this regard  --- and thus to guide experimenters on how to potentially go about detecting FQAH anyons ---
we highlight that, generally, the \emph{quantum adiabatic theorem} \cite{nlab:quantumAdiabaticTheorem} (which is well-known to explain the presence of anyons in ordinary FQH systems \cite{Arovas1984} and elsewhere \cite{Cheng2011}) 
implies (cf. \cite[\S2-3]{MySS2024}\cite[\S3]{SS23-EoS}) that topological ground states form \emph{local systems} or \emph{flat bundles} of Hilbert spaces over the \emph{classical parameter space} $P$ of the system, and thus, equivalently, linear representations of the fundamental group of the connected components of $P$ (cf. Fig. \ref{AdiabaticQuantumTransport}). 

One speaks of the system exhibiting \emph{topological order} \cite{nLab:topologicalorder} if the irreducible representations involved are higher dimensional. This is independent of whether adiabatic transport of the system is actually executed by an experimentalist; but it means that \emph{if} it is executed, then the topological ground states will be acted on by the given unitary operator, exhibiting a \emph{topological quantum gate}.

For generic anyonic models in position space, these parameters $p$ typically involve the configurations of anyon positions in the plane, whose fundamental groups are the \emph{braid group} $\mathrm{Br}_n$, making the system's ground states form a braid representation, reflecting its anyonic topological order (cf. \cite[\S3]{MySS2024}).

But for the crystalline insulators of interest here, the only available parameters $p$ are those that determine the crystal structure (such as the bindings in tight-binding models) --- and that is equivalently what is encoded by the Bloch Hamiltonians \eqref{EquivariantMapInIntroduction} for single electrons: The actual FQAH ground states, being those of many and strongly interacting such electrons, will depend on the single-electron Bloch Hamiltonian in a complicated way that (as is the case for most strongly-coupled quantum systems) is largely not understood --- but all that matters here is that it does continuously depend on it.
Because, by the above phenomenon (Fig. \ref{AdiabaticQuantumTransport}), such dependency means that for given topological phase (connected component of the parameter space), the topological quantum states form (unitary, finite-dimensional, irreducible) linear representations of the group of homotopy classes of loops in the space of Bloch Hamiltonians \eqref{EquivariantMapInIntroduction}, 
\begin{equation}
  \label{EquivariantMonodromyInIntroduction}
  \begin{tikzcd}[column sep=30pt]
    \BrillouinTorus
    \ar[
      in=34+90, out=-34+90,
      looseness=3.5,
      "{ G }"{description}
    ]
    \ar[
      rr, 
      bend left=30,
      phantom,
      "{}"{name=s, swap}
    ]
    \ar[
      rr, 
      bend right=30,
      phantom,
      shift left=3pt,
      "{}"{name=t}
    ]
    \ar[
      from=t, to=s,
      shorten=-4pt,
      Rightarrow,
      shift right=4pt,
      bend right=60,
      gray,
      scale=.5
    ]
    \ar[
      rr, 
      bend left=30,
      "{ H/\vert H \vert }",
    ]
    \ar[
      rr, 
      bend right=30,
      gray,
      shift left=3pt,
      crossing over
    ]
    \ar[
      from=s, to=t,
      shorten=-4pt,
      Rightarrow,
      shift right=4pt,
      bend right=60,
      crossing over
    ]
    &&
    S^2
    \mathrlap{\,,}
    \ar[
      in=34+90, out=-34+90,
      looseness=4.4,
      "{ G }"{description}
    ]
  \end{tikzcd}
\end{equation}
hence of the \emph{fundamental group} $\pi_1$ \cite[(A13)]{supp}
of the space of Bloch Hamiltonians (hence of valence band topologies), at the given basepoint $(C,c)$\eqref{GenericFormOfFragilePhases}:
\begin{equation}
  \label{MonodromyInIntroduction}
    \adjustbox{scale=0.8,
      margin=2pt,
      bgcolor=lightgray!70
    }{%
        \def\arraystretch{1}%
        \begin{tabular}{@{}c@{}}%
          \bf 
          Topological order:
          \\
          monodromy of ground
          \\
          state Hilbert space
        \end{tabular}%
      }%
  \;
  \HilbertSpace
  \in
  \pi_{\mathcolor{purple}{1}}
  \mathrm{Map}\big(
    \BrillouinTorus
    ,
    S^2
  \big)^{\!G}_{\!(C,c)}
  \mbox{-}
  \mathrm{Rep}\mathrlap{\,.}
\end{equation}

The relevance of expression \eqref{MonodromyInIntroduction} is highlighted by the remarkable fact
(due to \cite{SS25-FQAH,SS25-FQH,KSS26-HigherDimAnyons}, reviewed below in \S\ref{TopOrderForNoSymmetry}) that, when all symmetry is broken ($G = 1$),
it  gives exactly the anyon states as expected for FQH systems on a torus! --- but here: on the Brillouin torus of crystal momenta. In other words, basic (co)homotopy theory predicts clearly that the anyonic states of fractional Chern insulators, absent all symmetries, are localized not in position space but in reciprocal momentum space.

This motivates us to systematically analyze here the situation for non-trivial crystalline symmetries $G$. Curiously, we find that for all non-trivial $G$ the above solitonic anyons in momentum space disappear, but instead quantum states corresponding to \emph{defect} anyons subject to parastatistics appear for certain $G$, these defect anyons associated with the crystal's high symmetry points. 

To see this, we need to resolve a further subtlety that may not have received due attention before:

\subsection{Covariantization under Diffeos}

While the fragile topological phases according to \eqref{FragilePhasesIdeaInIntro} should be classically observable as such, it is another question whether the system's topological ground state Hilbert spaces \eqref{MonodromyInIntroduction} discriminate all these phases:

Namely, if the topological ground states follow the rules of topological quantum field theory (TQFT) and are hence ``generally covariant'' \cite{Witten1988} --- as is generally expected, cf. \cite{Simon2023} --- then any pair of band topologies differing only by a \emph{diffeomorphism} of the domain $\BrillouinTorus$ must be physically equivalent as seen by that ground state TQFT. 

Mathematically, such general covariance means (cf. \cite{Dul2023}) to pass to the \emph{stacky quotient} or \emph{homotopy quotient} ``$(\mbox{-})\!\sslash \! (\mbox{-})$'' 
\cite[(A33)]{supp}
of the space of band topologies by the (oriented) \emph{equivariant diffeomorphism group} $\mathrm{Diff}^+(\mbox{-})^{G}$ \cite{nlab:DiffeomorphismGroup}. 

For topological phases, hence on connected components, this simply means that would-be phases are identified via the ordinary quotient $(\mbox{-})/(\mbox{-})$ by the 
\emph{equivariant mapping class group} (for which cf. \cite{Cai2022}) of the torus, 
\begin{equation}
  \label{EquivariantMCG}
  \mathrm{Mod}^G
  :=
  \pi_0\Big( 
    \mathrm{Diff}^+\big(\BrillouinTorus\big)^{\!G}
  \Big)
  \mathrlap{\,,}
\end{equation}
(for short: the \emph{equivariant modular group}) in that:
\begin{widetext}
\begin{equation}
  \label{CovariantizedPhasesInIntro}
  \left\{
    \adjustbox{scale=0.8,
      margin=2pt,
      bgcolor=lightgray!70
    }{%
        \def\arraystretch{1}%
        \begin{tabular}{@{}c@{}}%
          \bf Fragile topological phases
          \\
          discernible by the system's 
          \\
          topological ground states
        \end{tabular}%
      }%
  \right\}
  =
  \grayunderbrace{
  \pi_0
  \Big(
    \mathrm{Map}\big(
      \BrillouinTorus
      ,\,
      S^2
    \big)^{\!G}
    \!\!\sslash
    \mathrm{Diff}^+\big(
      \BrillouinTorus
    \big)^{\!G}
  \Big)
  }{
    \scalebox{.7}{
      components of
      covariantized band topology
    }
  }
  \simeq
  \grayunderbrace{
  \Big(
  \pi_0 \, 
  \mathrm{Map}\big(
    \BrillouinTorus
    ,
    S^2
  \big)^{\!G}
  \Big)
  }{
    \scalebox{.7}{
      band topology components
    }
  }
  /
  \grayunderbrace{
  \mathclap{\phantom{\Big(}}
  \mathrm{Mod}^G
  }{
    \mathclap{
    \scalebox{.7}{
      \def\arraystretch{.9}
      \begin{tabular}{c}
        modular
        \\
        transformations
      \end{tabular}
    }
    }
  }
  \mathrlap{\,.}
\end{equation}
\end{widetext}

We will see that this does not affect the Chern numbers $C$ but does generally reduce the set of discrete phases $c$ in \eqref{GenericFormOfFragilePhases}.

More drastically, for the topological order \eqref{MonodromyInIntroduction}, hence on fundamental groups, this covariantization means
(cf. \cite[Prop. 2.24]{SS25-FQH}) 
that 
the monodromy of the band topology \eqref{EquivariantMonodromyInIntroduction} is \emph{extended} $(\mbox{-})\widehat{\times}(\mbox{-})$ by the  equivariant modular group \eqref{EquivariantMCG} :
\begin{widetext}
\begin{equation}
  \label{CovariantizedMonodromyInIntroduction}
    \adjustbox{scale=0.8,
      margin=2pt,
      bgcolor=lightgray!70
    }{%
        \def\arraystretch{1}%
        \begin{tabular}{@{}c@{}}%
          \bf Topological order
          \\
          reflecting covariance
          \\
          of top. Hilbert space
        \end{tabular}%
      }%
  \;
  \HilbertSpace
  \in
  \grayunderbrace{
  \pi_1 
  \Big(
    \mathrm{Map}\big(
      \BrillouinTorus
      ,
      S^2
    \big)^{\!G}
    \!\!\!
    \sslash
    \!
    \mathrm{Diff}^+\big(
      \BrillouinTorus
    \big)^{\!G}
  \Big)
  }{
    \mathclap{
    \scalebox{.7}{
      monodromy of covariantized band topology
    }
    }
  }
  \mbox{-}\mathrm{Rep}
  \begin{tikzcd}[
    column sep=15pt
  ]
    {}
    \ar[r]
    &
    {}
  \end{tikzcd}
  \grayunderbrace{
  \Big(
  \pi_1\,
  \mathrm{Map}\big(
    \BrillouinTorus
    ,\,
    S^2
  \big)^{\!G}
  \Big)
  }{
    \scalebox{.7}{
      band topology monodromy
    }
  }
  \widehat{\times}
  \grayunderbrace{
    \mathclap{\phantom{\Big(}}%
    \mathrm{Mod}^G
  }{
    \mathclap{
    \scalebox{.7}{
      \def\arraystretch{.9}
      \begin{tabular}{c}
        equivariant
        \\
        mapping classes
      \end{tabular}
    }
    }
  }
  \mbox{-}\mathrm{Rep}
  \mathrlap{\,.}
\end{equation}
\end{widetext}

In a key example discussed in \S\ref{OnTopologicalOrderForp3Symmetry}, of crystal symmetry $\SpaceGroup{p3}$, the $\ZThree$-modular group turns out to be the symmetric group, $\mathrm{Mod}^{\ZThree} \simeq \mathrm{Sym}_3$, generated by lattice translations and by rotations that permute the three high-symmetry points of crystal momenta among each other. With \eqref{CovariantizedMonodromyInIntroduction} this means that the topological quantum states in this case form a permutation representation, hence a ``very stable'' representation of the braid group $\mathrm{Br}_3$ (factoring through the surjection $\mathrm{Br}_3 \twoheadrightarrow \mathrm{Sym}_3$) and as such behave like quantum states of ``defect para-anyons'' localized at the high-symmetry points in momentum space.

\medskip

The expressions 
\eqref{FragilePhasesIdeaInIntro} and \eqref{MonodromyInIntroduction}
should be uncontroversial ---
while \eqref{CovariantizedPhasesInIntro} and \eqref{CovariantizedMonodromyInIntroduction} should be at least plausible ---
and are certainly foundational for the quantum topology of (fractional) Chern insulators --- yet it appears that none of them have been computed nor even considered before. At the same time, also the methods of equivariant (co)homotopy needed for their computation, while standard in the pure math literature, happen to be outside the topological toolbox that the condensed matter community has become acquainted with --- and so we aim to work towards filling this gap. Our supplementary material \cite{supp} provides details on computations, and its appendix \cite[\S A]{supp} compiles relevant background on homotopy theory.

\section{Introduction and Concepts}
\label{MotivationAndIntroduction}

We now explain the agenda of \S\ref{Agenda} in more detail.

Crystalline quantum materials called \emph{Chern insulators} (CI, \cite{AndoFu2015}\cite[\S8-9]{Sergeev2023}) famously exhibit ``topological phases'' (cf. \cite[\S II]{Stanescu2020}), due to the phenomenon that the Bloch quantum states \cite[\S XIII.16]{ReedSimon1978:IV}
of these crystals' valence electrons span --- as the electron momenta $\CrystalMomentum$ vary in the \emph{Brillouin torus} $\GeneralBrillouinTorus{d}$ of crystal momenta (cf. \cite[\S 2.1]{Thiang2025}) --- a nontrivial complex vector bundle $\mathcal{V}$:

\begin{equation}
  \label{GeneralValenceBundle}
  \begin{tikzcd}[
    column sep=12pt,
    row sep=4pt
  ]
    \smash{\mathllap{
      \substack{
        \acomment
          {Hilbert space of}
          {valence electrons}
          {at momentum $\CrystalMomentum$}
      }
      \;\;\;
    }}
    \ValenceBundle
      _{\scalebox{.7}{$\CrystalMomentum$}}
    \ar[
      ddr,
      phantom,
      "{
        \scalebox{.6}{\color{gray}(pb)}
      }"{pos=.2}
    ]
    \ar[
      dd
    ]
    \ar[
      rr,
      hook
    ]
    &&
    \ValenceBundle
    \smash{\mathrlap{
      \;\;\;\;
      \substack{
        \acomment
          {valence}
          {bundle}
      }
      \;\;\;
    }}
    \ar[
      dd
    ]
    \\
    \\
    \smash{\mathllap{
      \substack{
        \acomment
          {given}
          {momentum}
      }
      \;\;\;
    }}
    \big\{\CrystalMomentum\big\}
    \ar[
      rr,
      hook
    ]
    &
    {}
    &
    \GeneralBrillouinTorus{d}
    \smash{\mathrlap{
      \;\;
      \substack{
        \acomment
          {Brillouin torus of}
          {crystal momenta}
      }
    }}
  \end{tikzcd}
\end{equation}

Here we are concerned with the prominent case of effectively 2-dimensional CIs with a 2-band Bloch Hamiltonian (such as the seminal \emph{Haldane model}, cf. \cite[\S 8.3]{Sergeev2023}). Therefore the \emph{Bloch Hamiltonian} (cf. \cite[\S XIII.16]{ReedSimon1978:IV}\cite[\S5.1.3]{Sergeev2023}) may be expanded in the Pauli matrices $\{\sigma_i\}_{i=1}^3$ as:
\begin{equation}
  \label{TheBlochHamiltonian}
  \begin{tikzcd}[sep=0pt]
    \BrillouinTorus
    \ar[
      rr,
      "{
        H_{\mathrm{Blch}}
      }"
    ]
    &&
    \mathrm{Mat}_{2\times 2}(\mathbb{C})
    &[-5pt]
    \\
    \CrystalMomentum 
      &\longmapsto&
    H_{\mathrm{Blch}}(\CrystalMomentum)
    \\
    && \defneq\,
    h_0\big(\CrystalMomentum\big)
    &
    +
    \grayunderbrace{
      \textstyle{\sum_{i=1}^3}
      h_i\big(\CrystalMomentum\big)
      \, \sigma_i
    }{
      H(\CrystalMomentum)
    }
  \end{tikzcd}
\end{equation}
(for real coefficient functions $h$) and is \emph{gapped} in that its two energy eigenvalues never coincide:
\begin{equation}
  \big\vert H(\CrystalMomentum\,) \big\vert
  :=
  \sqrt{
  \textstyle{\sum_{i=1}^3}
  \big(
    h_i(\CrystalMomentum\,)
  \big)^2 
  }
  \;\;
    > 
  \;\;
  0
  \,,
\end{equation}
whence the normalization of the relative Bloch Hamiltonian $H$ in \eqref{TheBlochHamiltonian} constitutes a continuous map 
\begin{equation}
  \label{TheClassifyingMap}
  \vec\ClassifyingMap
  :=
  \vec h / {\vert H \vert}
\end{equation}
from $\BrillouinTorus$ to a 2-sphere $S^2$:
\begin{equation}
  \label{ClassifyingMap}
  \begin{tikzcd}[row sep=-3pt, column sep=5pt]
    \BrillouinTorus
    \ar[
      rr,
      "{ \ClassifyingMap }"
    ]
    &&
    S^2%
    \subset \mathbb{R}^3
    \\
    \CrystalMomentum
    &\longmapsto&
    \vec d\,(\CrystalMomentum\,)
    \mathrlap{\,.}
  \end{tikzcd}
\end{equation}
The valence bundle $\mathcal{V}$ 
\eqref{GeneralValenceBundle}
of lower eigenstates of \eqref{TheBlochHamiltonian}
is then the \emph{pullback} (pb) \cite[(A1)]{supp}
of the tautological/universal complex line bundle along this classifying map \eqref{ClassifyingMap}:
\begin{equation}
  \label{TheUniversalLineBundle}
  \hspace{-10mm} 
  \begin{tikzcd}[
    column sep=18pt,
    row sep=-1pt
  ]
    \mathclap{
      \substack{
        \acomment
          {valence}
          {bundle}
      }
    }
    & &[-5pt] 
    \mathclap{
      \substack{
        \acomment
          {tautological complex}
          {line bundle}
      }
    }
    & &[-5pt]
    \mathclap{
     \substack{
       \acomment
         {universal complex}
         {line bundle}
     }
    }
    \\
    \ValenceBundle
    \ar[
      rr
    ]
    \ar[d]
    \ar[
      dr,
      phantom,
      "{
        \scalebox{.6}{\color{gray}(pb)}
      }"{pos=.2}
    ]
    &&
    \mathcal{L}
    \ar[d]
    \ar[rr]
    \ar[
      dr,
      phantom,
      "{
        \scalebox{.6}{\color{gray}(pb)}
      }"{pos=.2}
    ]
    &&
    \smash{
    E\mathrm{U}(1)
      \mathrlap{
      \,
      \underset{\mathclap{\mathrm{U}(1)}}{\times}
      \,
      \mathbb{C} 
    }
    }
    \ar[
      d
    ]
    \\[+15pt]
    \BrillouinTorus
    \ar[
      r, 
      "{ \ClassifyingMap }"
    ]
    &
    S^2
    \ar[
      r,
      equals
    ]
    &
    \mathbb{C}P^1
    \ar[
      r,
      hook
    ]
    &
    \mathbb{C}P^\infty
    \ar[
      r,
      "{ \sim }"{pos=.35}
    ]
    &
    B \mathrm{U}(1)
    \mathrlap{\,,}
    \\[-3pt]
    &&
    \mathclap{
     \substack{
       \acomment
         {fragile}
         {band topology}
     }
    }
    &
    {}
    \ar[
      r,
      phantom,
      "{
        \mathclap{
          \substack{
           \acomment
             {stable}
             {band topology}
          }
        }
        }"
    ]
    &
    {}
  \end{tikzcd}
\end{equation}
hence is equivalently classified by the homotopy class $\big[\ClassifyingMap\big]$
\cite[A9]{supp}
of $d$ --- which in this case is the \emph{Hopf degree} (or \emph{winding number}) coinciding with the (first and only) \emph{Chern number} of the valence bundle:
\begin{equation}
  \label{TheWindingNumber}
  C 
  :=
  \big[\ClassifyingMap\big]
  \;\in\;
  \grayunderbrace{
  \mathcolor{darkorange}{\pi_0}
  \,
  \mathrm{Map}\big(
    \BrillouinTorus
    ,\,
    S^2
  \big)
  }{
    \pi^2(\BrillouinTorus)
  }
  \;\simeq\;
  \mathbb{Z}
  \,.
\end{equation}
As shown under the brace, it is worth highlighting here that \eqref{TheWindingNumber} is the \emph{2-Cohomotopy} of the torus \cite[\S 3]{SS25-FQH}, where \emph{$n$-Cohomotopy} $\pi^n(-)$ \cite[\S VII]{STHu59}\cite[p. 31]{SS25-Flux}
is the ``extraordinary'' (or ``generalized'') and ``non-abelian'' (or ``unstable'') cohomology theory \cite[\S 2]{FSS23-Char} which is represented by the $n$-sphere:
\begin{equation}
  \label{nCohomotopy}
  \pi^n(\mbox{-})
  :=
  \pi_0\, 
  \mathrm{Map}(\mbox{-},S^n)
  \mathrlap{\,.}
\end{equation}
This is analogous to how ordinary integral cohomology is classified by \emph{Eilenberg-MacLane spaces} $B^n \mathbb{Z} \defneq K(\mathbb{Z},n)$ \cite[Ex. 2.1]{FSS23-Char}
\begin{equation}
  H^n(-;\mathbb{Z})
  \simeq
  \pi_0\,
  \mathrm{Map}\big(-, B^n \mathbb{Z}\big)
\end{equation}
and how K-cohomology is classified by spaces $\mathrm{Fred}$ of Fredholm operators (cf. \cite{Atiyah1969,SS25-OrbiK}):
\begin{equation}
  K^n(-)
  \simeq
  \pi_0\,
  \mathrm{Map}\big(-, \Sigma^n \mathrm{Fred}\big)
  \,.
\end{equation}
Here, for topological phases on the plain 2-torus, with $n = 2$, all of these cohomology theories agree. But we will see that this coincidence is lifted as more structure is taken into account, and that the faithful description of the fragile band topology by 2-Cohomotopy stands out.

Namely, to the above standard picture (cf. \cite{Sergeev2023}) we now add --- following \cite{SS25-FQAH} --- three fundamental observations which appear to have escaped attention before: 

\begin{itemize}
  \item[\S\ref{SymmetryAndEquivariance}:]
  \emph{crystalline $G$-symmetry protection} means that the classifying map \eqref{ClassifyingMap} is constrained to be \emph{$G$-equivariant}, hence that the space of fragile band topologies is the \emph{equivariant} mapping space $\mathrm{Map}\big(\BrillouinTorus, S^2\big)^{\!G}$,

  \item[\S\ref{AnyonsAndMonodromy}:]
  \emph{topological order} means that gapped ground states transform under \emph{monodromy} in this space of band topologies, hence form linear representations of its fundamental groups $\pi_1$,

  \item[\S\ref{CovarianceUnderDiffeomorphisms}:] 
  \emph{topological quantum} phenomenology furthermore means that these 
  topological quantum states must be \emph{generally covariant} with respect to (equivariant) diffeomorphisms, whence the covariantized space of band topologies as seen by the system's own ground state TQFT is really the homotopy quotient $\mathrm{Map}\big(\BrillouinTorus, S^2\big)^{\!G} \!\sslash\! \mathrm{Diff}^+\big(\BrillouinTorus\big)^{\!G}$.
\end{itemize}

We expand on these three statements in turn:

\subsection{Symmetry and Equivariance}
\label{SymmetryAndEquivariance}

It is now commonplace in the theory of topological phases that \emph{$G$-crystalline symmetry} means (cf. \cite{NeupertSchindler2018}\cite[\S 5.2]{Stanescu2020}):

\begin{enumerate}
\item[\bf (i)] A symmetry group $G$ acting on the crystal lattice and hence dually on the Brillouin torus $\GeneralBrillouinTorus{d}$
\begin{equation}
  \begin{tikzcd}[row sep=-3pt, column sep=0pt]
  G \times \GeneralBrillouinTorus{d}
  \ar[
    rr
  ]
  &&
  \GeneralBrillouinTorus{d}
  \\
  \big(
    g, 
    \CrystalMomentum
  \big)
  &\longmapsto&
  g \cdot \CrystalMomentum
  \mathrlap{\,;}
  \end{tikzcd}
\end{equation}

\item[\bf (ii)] a compatible family of unitary operators (on $n$ bands, say)
\begin{equation}
  \label{TheUnitaryOperators}
  \begin{tikzcd}[row sep=-3pt, column sep=0pt]
    G \times \GeneralBrillouinTorus{d}
    \ar[
      rr,
      "{ U }"
    ]
    &&
    \mathrm{U}(n)
    \\
    \big(g,\, \CrystalMomentum \big)
    &\longmapsto&
    U_g\big(\CrystalMomentum \big)
  \end{tikzcd}
\end{equation}
\item[\bf (iii)]  such that the Bloch Hamiltonian satisfies this condition
\begin{equation}
  \label{TransformationOfHamiltonian}
  \begin{aligned}
  &
  \grayoverbrace{
  H_{\mathrm{Blch}}\big(
    g \cdot \CrystalMomentum
   \big)
  }{%
    \mathclap{%
     \substack{%
       \acomment%
         {crystalline symmetry}
         {transform}%
     }%
    }%
  }
  \\
  &
  =
  \grayunderbrace{
  U_g(\CrystalMomentum)
    \,\circ\, 
  H_{\mathrm{Blch}}(\CrystalMomentum)
    \,\circ\,
  U_{g}(\CrystalMomentum)^{-1}
  }{
   \substack{
     \acomment
       {band symmetry transform}
   }
  }
  \,.
  \end{aligned}
\end{equation}
for all $g \in G$ and $\CrystalMomentum \in \GeneralBrillouinTorus{d}$.
\end{enumerate}
We highlight now that this condition 
\eqref{TransformationOfHamiltonian}
makes the Bloch Hamiltonian an \emph{orbifold} map (cf. \cite{SS26-Orb}), and in the common special case that the unitary operators \eqref{TheUnitaryOperators} are actually independent of $\CrystalMomentum$, 
\begin{equation}
  \label{RestrictedTransformationOfHamiltonian}
  H_{\mathrm{Blch}}\big(
    g \cdot \CrystalMomentum
  \big)
  =
  U_g
    \circ 
  H_{\mathrm{Blch}}\big(\CrystalMomentum\big)
    \circ
  U_{g}^{-1}
  \mathrlap{\,,}
\end{equation}
it says equivalently that the Bloch Hamiltonian is a \emph{$G$-equivariant map} \cite[(A39)]{supp},
\begin{equation}
  \label{BlochHamiltonianAsEquivariantMap}
  \begin{tikzcd}[
    sep=0pt
  ]
    \GeneralBrillouinTorus{d}
    \ar[
      in=34+90, out=-34+90,
      looseness=3.2,
      "{ G }"{description}
    ]
    \ar[
      rr
    ]
    &&
    \mathrm{Mat}_{n \times n}(\mathbb{C})
    \ar[
      in=34+90, out=-34+90,
      looseness=4.4,
      "{ G }"{description}
    ]
    \\
    \CrystalMomentum
    &\longmapsto&
    H_{\mathrm{Blch}}\big(\CrystalMomentum\big)
    \mathrlap{\,,}
  \end{tikzcd}
\end{equation}
with respect to the conjugation action of $G$ on the right, through the homomorphism
\begin{equation}
  \begin{tikzcd}
   G 
   \ar[r, "{ U }"]
   &
   \mathrm{U}(n)
   \ar[r, ->>]
   &
   \mathrm{PU}(n)
   \;\acts\; 
   \mathrm{Mat}_{n \times n}(\mathbb{C})
   \mathrlap{\,.}
  \end{tikzcd}
\end{equation}

Specifically, in the case at hand of $n=2$-band Bloch Hamiltonians \eqref{TheBlochHamiltonian},
crystalline $G$-symmetry \eqref{BlochHamiltonianAsEquivariantMap} means that the classifying map \eqref{ClassifyingMap} is $G$-equivariant
\begin{equation}
  \label{EquivariantClassifyingMap}
  \begin{tikzcd}[
    sep=20pt
  ]
    \BrillouinTorus
    \ar[
      in=34+90, out=-34+90,
      looseness=3.5,
      "{ G }"{description}
    ]
    \ar[
      rr,
      "{ \ClassifyingMap }"
    ]
    &&
    S^2
    \ar[
      in=34+90, out=-34+90,
      looseness=4.4,
      "{ G }"{description}
    ]
  \end{tikzcd}
\end{equation}
with respect to, on the right, the rotation action
\begin{equation}
  \label{GRotationActionOnS2}
  \begin{tikzcd}
    G 
    \ar[r, "{ U }"]
    &
    \mathrm{U}(2)
    \ar[r]
    &
    \mathrm{PU}(2)
    \simeq
    \mathrm{SO}(3)
    \;\acts\;
    S^2
    \mathrlap{\,.}
  \end{tikzcd}
\end{equation}

Now, \emph{$G$-symmetry protection} 
of topological phases \cite{nLab:SPT} means that with the Bloch Hamiltonian itself also all its available deformations satisfy the condition \eqref{RestrictedTransformationOfHamiltonian}, hence that the space of available band topologies is now the subspace
of $G$-equivariant maps \eqref{EquivariantClassifyingMap} among all maps \cite[(A41)]{supp}:
\begin{equation}
  \label{SymmetricFragileBandTopologySpace}
  \begin{tikzcd}[sep=0pt]
  \mathrm{Map}\big(
    \BrillouinTorus
    ,
    S^2
  \big)^{\!G}
  &\subset&
  \mathrm{Map}\big(
    \BrillouinTorus
    ,
    S^2
  \big)
  \mathrlap{\,.}
  \\
  \substack{
    \acomment
      {$G$-symmetric}
      {fragile band topologies}
  }
  &&
  \substack{
    \acomment
      {fragile}
      {band topologies}
  }
  \end{tikzcd}
\end{equation}

This equivalent reformulation of $G$-symmetry protection of topological phases is noteworthy, because there is a well-developed and sophisticated mathematical machinery for analyzing the invariant content of equivariant mapping spaces like \eqref{SymmetricFragileBandTopologySpace}: this is \emph{equivariant homotopy theory} (cf. \cite{nLab:EquivariantHomotopyTheory}\cite[\S A 2]{supp}).

Concretely, where the connected components on the right of \eqref{SymmetricFragileBandTopologySpace}
give the \emph{2-Cohomotopy} of the Brillouin torus \eqref{TheWindingNumber},  the connected components on the left of \eqref{SymmetricFragileBandTopologySpace} constitute the \emph{equivariant 2-Cohomotopy} \cite[Ex. 4.5.8]{SS25-Bun}\cite[\S 6]{SS26-Orb}\cite[(3)]{SS20-Tad}\cite{BSS21-Frac} of the Brillouin torus \eqref{FragilePhasesIdeaInIntro}:
\begin{equation}
  \label{Equivariant2Cohomotopy}
  \pi_G^2\big(
    \BrillouinTorus
  \big)
  \defneq
  \pi_0\, 
  \mathrm{Map}\big(
    \BrillouinTorus
    ,
    S^2
  \big)^{\!G}
  \mathrlap{\,,}
\end{equation}
and the prediction is hence that 
this equivariant 2-Cohomotopy classifies the $G$-symmetry protected fragile band topology of 2D 2-band Chern insulators. 

This is the first set of quantities that we will report on, in \S\ref{TopologicalPhases}.

\subsection{Topological Order and Monodromy}
\label{AnyonsAndMonodromy}

\paragraph{\bf \small FQAH Systems.}

Certain Chern insulators at fractional filling of their ``flat'' valence band --- \emph{fractional Chern insulators} (FCI) --- had been predicted \cite{Neupert2011}, and have just recently been experimentally observed \cite{Cai2023, Zeng2023, Park2023, Lu2024}, to exhibit an intrinsic form of what for 2D electron gases in strong transversal magnetic fields is the \emph{fractional quantum Hall} effect (FQH, cf \cite{nLab:FQH}), hence here called the \emph{fractional quantum anomalous Hall effect} (FQAH). 

The adjective ``anomalous'' here is a historical expression of the surprise that a magnetic field, which is so paramount for the ordinary Quantum Hall effect, does not play a role in the anomalous version. Instead, it is the \emph{Berry curvature} of the valence band over the Brillouin torus of crystal momenta which plays the role of magnetic flux for FQAH systems (cf. \cite[Fig. 3]{SS25-FQAH}). 

\begin{figure}[htb]
\caption{
\label{PositionMomentumDuality}
  There is a duality between the FQH effect and its anomalous FQAH version, under which the ordinary space of positions inside the 2D material is exchanged for the ``reciprocal'' space of crystal momenta, while the external magnetic flux density is exchanged for the Berry curvature of the Bloch bands over this momentum space.
}

\centering

\adjustbox{
  scale=0.9,
  raise=-1.6cm,
  rndfbox=4pt
}{
\begin{tikzpicture}

\node[scale=.7] at (-2.9,1.2) {%
  \color{gray}%
  \bf%
  \def\arraystretch{.9}%
  \begin{tabular}{@{}c@{}}%
    Semiclassical 
    \\
    equation of motion
    \\
    of crystal electrons
  \end{tabular}
};

\node at (0,0) {
\begin{tikzcd}[
  column sep=-2pt,
  row sep=10pt
]
  \dot {\vec k}
  &=&
  -
  \partial \epsilon
  /
  \partial\vec x
  &+&
  \overset{
    \mathclap{
      \adjustbox{
        scale=.7,
        raise=2pt
      }{
        \color{darkblue}
        \bf
        Lorentz force
      }
    }
  }{
    \vec x \,\times\, e\vec B
  }
  &\phantom{---}&
  \overset{
    \mathclap{
      \adjustbox{
        scale=.7,
        raise=4pt
      }{
        \color{darkblue}
        \bf
        \def\arraystretch{.9}
        \begin{tabular}{c}
          position
        \end{tabular}
      }
    }
  }{
    \vec x
  }
  \ar[
    <->,
    shorten=-2pt,
    dashed,
    gray,
    d
  ]
  &\phantom{---}&
  \overset{%
    {%
      \adjustbox{
        scale=.7,
      }{%
        \color{darkblue}%
        \bf%
        \def\arraystretch{.9}%
        \begin{tabular}{@{}c@{}}%
          magnetic
          \\
          flux density
        \end{tabular}%
      }%
    }%
  }{
    e \vec B
  }
  \ar[
    <->,
    shorten=-2pt,
    dashed,
    gray,
    d
  ]
  \\
  \dot {\vec x}
  &=&
  \phantom{+}
  \partial \epsilon
  /
  \partial\vec k
  &-&
  \underset{
    \mathclap{
      \adjustbox{
        scale=.7,
        raise=-2pt
      }{
        \color{darkblue}
        \bf
        anomalous velocity
      }
    }  
  }{
    \vec k \,\times\, \phantom{e}\vec\Omega
  }
  &&
  \underset{
    \mathclap{
      \adjustbox{
        scale=.7,
        raise=-3pt
      }{
        \color{darkblue}
        \bf
        \def\arraystretch{.9}
        \begin{tabular}{c}
          momentum
        \end{tabular}
      }
    }
  }{
    \vec k
  }
  &&
  \underset{
    \mathclap{
      \adjustbox{
        scale=.7,
        raise=-6.5pt
      }{
        \color{darkblue}
        \bf
        \def\arraystretch{.85}
        \begin{tabular}{c}
          Berry 
          \\
          curvature
        \end{tabular}
      }
    }
  }{
    \vec \Omega
  }
\end{tikzcd}
};
\begin{scope}[
  shift={(-2.3,0)}
]
\draw[
  <->,
  densely dashed
]
  (2.25,.5) .. controls
  (3.2,.5) and
  (3.2,-.5) ..
  (2.25,-.5);
\end{scope}

\node[scale=.8] at (2.6,1.5) {
  \color{darkorange}
  \bf
  Hall effect
};
\node[scale=.8] at (2.8,-1.5) {
  \color{darkorange}
  \bf
  \llap{Anomalous }Hall effect
};

\end{tikzpicture}
}
\end{figure}
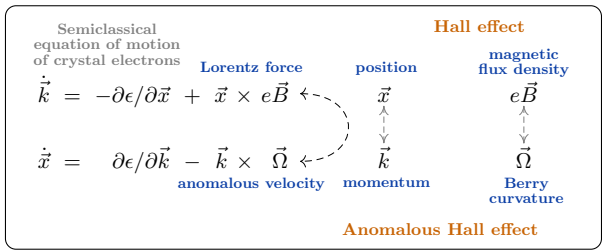

Therefore, FQAH systems are not actually a freakish anomaly in FQH phenomenology, as the terminology may suggest, but rather are systematic \emph{duals} to FQH systems (in the sense of duality familiar in high energy physics), under a duality which swaps ordinary position space with momentum space, and magnetic flux with Berry curvature
(an observation that goes back to \cite{Chang1995} \cite{Chang1996}\cite[\S III.B]{Sundaram1999}, 
review in \cite{Xiao2005}\cite[(69)]{Sinitsyn2007}\cite[\S III.A]{Xiao2010}\cite{Chong2010}\cite[(12-13)]{Stephanov2012}), see Fig. \ref{PositionMomentumDuality}.

But since ordinary FQH systems are famously known to exhibit anyonic topological order --- experimentally observed for abelian anyons 
since \cite{Nakamura2020}, recent pointers in \cite{Veillon2024} --- it is tantalizingly suggestive that so should their ``anomalous'' cousins, in some accordingly dual way. If so, this would potentially have dramatic technological consequences, as FQAH systems are realized under much more practical conditions than FQH systems (not requiring their strong magnetic fields and low temperatures: whence the prospect of ``high-temperature topological order'' \cite{Parameswaran2013}). Therefore, understanding anyonic topological order in FQAH systems could potentially pave the way to practical topological quantum hardware for much anticipated robust quantum computers.

However, there has been no theoretical prediction (in fact no consideration, it appears) for how anyonic topological order could actually manifest in FQAH systems, and hence no guide to experiment for how to go about detecting it.

\paragraph{\bf \small General Topological Order.}

In order to address this open problem, we now step back and revisit the general question:

\begin{standout}
\emph{What is the manifestation of topological quantum order, generally?}
\end{standout}
Namely, the now widely popular idea that topological order \cite{nLab:topologicalorder}
and topological quantum processes (as in: topological quantum gates) are reflected by (degenerate quantum ground states acted on) by braiding processes of worldlines of point particles (anyons) is \emph{only one instance} of a general concept of topological quantum processes governed by the \emph{quantum adiabatic theorem} \cite{nlab:quantumAdiabaticTheorem}, which, we highlight, in general only requires the following three \emph{conditions for topological quantum order} (cf. \cite[\S 2-3]{MySS2024}):

\begin{definition}[General topological order]
\label{GeneralTopologicalOrder}
A quantum system exhibits (\emph{proper}) \textbf{topological order} if the following three conditions are satisfied:
\begin{enumerate}
  \label{ThreeConditions}
  \item
  \label{FirstCondition}
  \textbf{(Holonomy)}
  The Hilbert spaces $\HilbertSpace_p$ of gapped quantum ground states  \emph{depend} differentiably on classical external parameters $p$ varying in some parameter space $P$.

  By the quantum adiabatic theorem \cite{nlab:quantumAdiabaticTheorem}, this implies that adiabatic tuning (meaning: slow with respect to the quantum system's relaxation time) of the parameters along continuous paths $\!\begin{tikzcd}[sep=21pt, decoration=snake] p \ar[r, decorate, "{\gamma\;}"] & p' \end{tikzcd}\!$ in $P$ induces (asymptotically) unitary transformations $U_\gamma$ of the quantum system:
  \begin{equation}
    \label{UnitaryTransformationOverPath}
    \begin{tikzcd}[
      row sep=-3pt, column sep=15pt, 
      decoration=snake
    ]
      \HilbertSpace_p
      \ar[
        rr,
        "{ U_\gamma }"
      ]
      &&
      \HilbertSpace_{p'}
      \\
      p
      \ar[
        rr,
        decorate,
        "{ \gamma }"
      ]
      &&
      p'
      \mathrlap{\,.}
    \end{tikzcd}
  \end{equation}
  
  \item 
  \label{SecondCondition}
  \textbf{(Homotopy)}
  This dependency on parameter paths is ``topological'' (which is physics jargon for what in mathematics really is: \emph{homotopical}) in that a continuous deformation $\sigma : \gamma \Rightarrow \gamma'$ of parameter paths (\emph{homotopies}, keeping their endpoints fixed) does not affect the induced unitary transformations:
  \begin{equation}
    \begin{tikzcd}[
      row sep=2pt,
      decoration=snake
    ]
      \HilbertSpace_p
      \ar[
        rr,
        "{ 
          U_\gamma 
          \,=\,
          U_{\gamma'}
        }"
      ]
      &&
      \HilbertSpace_{p'}
      \\
      p
      \ar[
        rr,
        decorate,
        bend left=10,
        "{ \gamma }"{
          description,
          name=s
        }
      ]
      \ar[
        rr,
        decorate,
        bend right=40,
        "{ \gamma' }"{
          description,
          name=t
        }
      ]
      \ar[
        from=s,
        to=t,
        Rightarrow
      ]
      &&
      p'
      \mathrlap{\,.}
    \end{tikzcd}
  \end{equation}

  \item
  \label{ThirdCondition}
  \textbf{(Monodromy)}
  The fundamental group $\pi_1$ (of any connected component) of $P$ is non-trivial (properly: non-abelian), in that some such pairs of parameter paths with the same endpoints do \emph{not admit} (\#) continuous deformations into each other,   \begin{equation}
    \begin{tikzcd}[
      row sep=5pt,
      decoration=snake
    ]
      \HilbertSpace_p
      \ar[
        rr,
        bend left=15,
        "{ U_\gamma }"
      ]
      \ar[
        rr,
        bend right=30,
        "{ U_{\gamma'} }"
      ]
      &&
      \HilbertSpace_{p'}
      \\
      p
      \ar[
        rr,
        decorate,
        bend left=10,
        "{ \gamma }"{
          description,
          name=s
        }
      ]
      \ar[
        rr,
        decorate,
        bend right=40,
        "{ \gamma' }"{
          description,
          name=t
        }
      ]
      \ar[
        from=s,
        to=t,
        phantom,
        "{ \# }"
      ]
      &&
      p'
      \mathrlap{\,,}
    \end{tikzcd}
  \end{equation}
  and the induced unitary representation of the fundamental group 
\begin{equation}
  \begin{tikzcd}[
    row sep=-2pt, column sep=0pt,
    decoration=snake
  ]
    \pi_1(P, p_0)
    \ar[rr]
    &&
    \mathrm{U}\big(
      \HilbertSpace_{p_0}
    \big)
    \\
    p_0
    \ar[
      in=40-90,
      out=-40-90,
      looseness=4.7,
      shift left=2pt,
      decorate,
      "{ \gamma }"
    ]
      &\longmapsto& 
    \HilbertSpace_{p_0}
    \ar[
      in=41-90,
      out=-41-90,
      shift left=3pt,
      looseness=5,
      "{ 
        U_{\gamma}\! 
      }"{description}
    ]
  \end{tikzcd}
\end{equation}  
  is non-trivial (properly: contains irreps of dimension $>1$).
\end{enumerate}

\end{definition}

Physically:
\begin{itemize}
\item The \ref{FirstCondition}st condition says that tuning of parameters induces transformations of gapped quantum ground states like \emph{holonomic quantum gates},
\item the \ref{SecondCondition}nd condition says that this operation is ``topological'' (really: \emph{homotopical}) in that it is insensitive to local deformation (noise!) in the parameter paths, 
\item and the \ref{ThirdCondition}rd condition ensures that there are non-trivial such topological processes inducing non-trivial transformations on the quantum system. 
\end{itemize}
We note that no experimentalist needs to execute such topological processes for the system's ground states to represent them, but executing them amounts to applying \emph{topological quantum gates} to the system.

In mathematical jargon: 
\begin{itemize}
\item
the \ref{FirstCondition}st condition makes 
(cf. \cite{SchreiberWaldorf2009})
the Hilbert spaces $\HilbertSpace$ form a \emph{bundle with connection} over $P$ whose \emph{parallel transport} are the unitaries $\gamma \mapsto U_\gamma$, 

\item the \ref{SecondCondition}nd condition says that this connection is \emph{flat}, making the bundle of Hilbert spaces a \emph{local system} (cf. \cite[Lit 2.22]{MySS2024}\cite[\S3]{SS23-EoS}), 

\item the \ref{ThirdCondition}rd condition says that this local system is non-trivial (properly: nonabelian).
\end{itemize}

We next consider the familiar examples of this general notion for the case of FQH systems, and then its novel specialization to FQAH system.

\begin{example}[Traditional anyon braiding]
For plain FQH systems \cite{nLab:FQH}
on a surface $\Sigma^2$, the parameter space is that of configurations $S \subset \Sigma^2$ of quasi-hole/vortex positions, known as the surface's \emph{configuration space of points}  (cf. \cite[Lit. 2.18]{MySS2024}\cite{Kallel2025}):
\begin{equation}
  \label{ConfigurationSpaceOfPoints}
  P
  \defneq
  \mathrm{Conf}(\Sigma^2)
  :=
  \big\{
    S \subset \Sigma^2
    \,\big\vert\,
    S \in \mathrm{FinSet}
  \big\}
  \mathrlap{\,.}
\end{equation}
The fundamental groups of this parameter space \eqref{ConfigurationSpaceOfPoints}, in the connected component of $n$ vortices, are the surface \emph{braid groups}
(\cite{nLab:BraidGroup}, cf. \cite[\S2.2]{SS25-FQH})
\begin{equation}
  \pi_1\big(
    \mathrm{Conf}(\Sigma^2)
    ,
    n
  \big)
  \simeq
  \mathrm{Br}_n(\Sigma^2)
  \mathrlap{\,,}
\end{equation}
and their linear representations are the \emph{braid representations}, exhibiting ``braid group statistics'' of the anyonic vortices, cf. Fig. \ref{BraidRepresentationByTransport}.

\begin{figure}[htp]
\caption{
  \label{BraidRepresentationByTransport}
  The quantum adiabatic theorem implies that gapped quantum ground states depending ``topologically'' (homotopically) on the configurations of points in a surface, transform under unitary \emph{braid representations} as the configurations trace out loops in parameter space, forming \emph{braids} of worldlines.
}
\centering
\adjustbox{
  rndfbox=4pt
}{
\begin{tikzcd}[
  column sep=35
]
\HilbertSpace_6 
\ar[
  rr,
  "{ U_{\mathrm{braid}} }"
]
&&
\HilbertSpace_6
\\[15pt]
{}
\ar[
  rr, 
  phantom,
    "
    \adjustbox{
      scale={1}{1.2},
      rotate=-90
    }{
    \includegraphics[width=2cm]{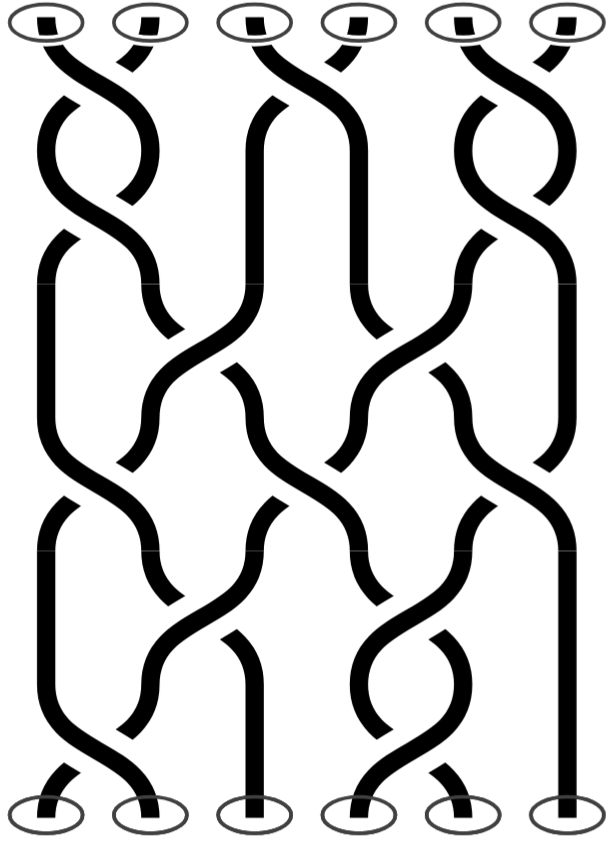}
    }
    "
]
\ar[
  rr,
  phantom,
  "{ 
    \mathrm{braid}
  }"{scale=.8, pos=.6}
]
&&
{}
\end{tikzcd}
}
\end{figure}
The theoretical identification of FQH anyons via the adiabatic quantum theorem applied to this situation goes back to \cite{Arovas1984, Su1986} and this is how (FQH) anyons have been conceptualized ever since, in particular in the literature on topological quantum computing (cf. \cite{Freedman2003, Nayak2008, Cheng2011, Cesare2015}).

Incidentally, flat bundles of Hilbert spaces over the parameter space \eqref{ConfigurationSpaceOfPoints} famously (but not exclusively) arise as \emph{Knizhnik-Zamolodchikov connections} \cite{nLab:KZConnection} on bundles of \emph{conformal blocks}, the resulting braid representations are often known (somewhat nondescriptively) as \emph{monodromy representations}.
\end{example}

But, contrary to the impression that one may get from perusal of the literature, braid representations are not the only way in which anyonic topological order may manifest in physical systems:

\begin{example}[Topological order over the torus]
\label{TraditionalTopologicalOrderOverTheTorus}
The phenomenology of the following example reproduces traditional lore about topological order often interpreted in terms of anyons --- but it does not involve anyon position parameters of the above kind \eqref{ConfigurationSpaceOfPoints}. We suggest that the way we obtain this now as another special case of Def. \ref{GeneralTopologicalOrder} speaks to the relevance of this definition.

Namely, it is commonly expected (cf. \cite[(4.21)]{Fradkin2013}\cite[(5.28)]{Tong2016})
that the algebra of topological quantum observables of an FQH system on the plain torus, $\Sigma^2 \defneq T^2$, is generated by a pair of unitary operators $W_a$, $W_b$ subject to the commutation relation (cf. Fig. \ref{TorusGroupCommutator})
\begin{equation}
  \label{TopologicalTorusObservables}
  W_a \, W_b
  =
  \zeta^2
  \,
  W_b W_a 
  \,,
\end{equation}
where $\zeta \in \mathbb{C}^\times$ is a root of unity identified with the FQH anyon braiding phase.

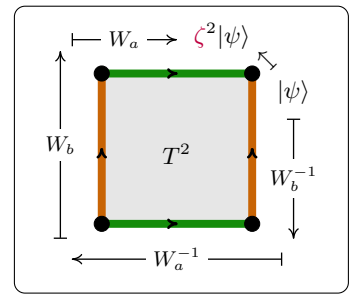
\begin{figure}[htbp]
\caption{%
  \label{TorusGroupCommutator}%
  The non-trivial group commutator in the algebra of anyon quantum observables on a torus in terms of the \emph{anyon braiding phase} $\zeta$. 
}
\centering
\adjustbox{
  rndfbox=4pt
}{
\begin{tikzpicture}[
  >={
    Computer Modern Rightarrow[
      length=4pt, width=4pt
    ]
  }
]

\draw[
  fill=lightgray,
]
  (0,0) rectangle (2,2);

\node at (1,.8) {%
  \clap{\smash{$T^2$}}%
};

\draw[
  line width=3,
  color=darkorange
] 
  (0,0) --
  (0,2);
\draw[
  ->,
  line width=1
]
  (0,1) --
  (0,1.01);

\draw[
  line width=3,
  color=darkorange
] 
  (2,0) --
  (2,2);
\draw[
  ->,
  line width=1
]
  (2,1) --
  (2,1.01);

\draw[
  line width=3,
  color=darkgreen
] 
  (0,2) --
  (2,2);
\draw[
  ->,
  line width=1
]
  (1,2) --
  (1.01,2);

\draw[
  line width=3,
  color=darkgreen
] 
  (0,0) --
  (2,0);
\draw[
  ->,
  line width=1
]
  (1,0) --
  (1.01,0);

\draw[
  fill=black
] 
(0,0) circle (.1);
\draw[
  fill=black
] 
(2,0) circle (.1);
\draw[
  fill=black
] 
(0,2) circle (.1);
(2,0) circle (.1);
\draw[
  fill=black
] 
(2,2) circle (.1);

\node at (2-.4,2+.5) {
  $\mathcolor{purple}{\zeta}^2\vert \psi \rangle$
};

\node[
  rotate=+135  
] at (2+.2, 2+.2) {%
  \clap{$\mapsto$}%
};

\node at (2.55,2-.25) {
  $\vert \psi \rangle$
};

\draw[
  line width=.5,
  |->,
]
  (2.55, 1.4) --
  node[
    xshift=0pt,
    scale=.9
  ] {\colorbox{white}{$
    W_b
      ^{-1}
  $}}
  (2.55,-.2);

\draw[
  line width=.5,
  |->,
]
  (2.4, -.47) --
  node[
    yshift=0pt,
    scale=.9
  ]{\colorbox{white}{$
    W_a^{-1}
  $}}
  (-.4, -.47);

\draw[
  line width=.5,
  |->,
]
  (-.55, -.2) --
  node[
    xshift=0pt,
    scale=.9
  ] {\colorbox{white}{$
    W_b
  $}}
  (-.55,2.3);

\draw[
  line width=.5,
  |->,
]
  (-.4, 2.45) --
  node[
    scale=.9,
  ]{\colorbox{white}{$
    W_a
  $}}
  (1, 2.45);
  
\end{tikzpicture}
}

\end{figure}

This topological observable algebra \eqref{TopologicalTorusObservables} is traditionally argued by appeal to an effective abelian Chern-Simons field theory (\cite{nLab:AbelianChernSimons}, cf. \cite[\S A.1]{SS25-FQH})
thought to describe the large-scale (``infrared'') dynamics of the FQH system. 

(Experimental verification of this expectation has been out of reach: While it might barely be possible to constrain a 2D electron gas to a torus, it seems nigh impossible to then have it penetrated by an everywhere transverse magnetic field. This makes it all the more remarkable that, as we pass to FQAH systems, the torus topology becomes not only experimentally viable but the default --- now for a \emph{momentum space} torus --- and our method for deriving FQH topological order on the torus seamlessly carries over to the FQAH situation.)

So we are to ask:
\begin{standout}
Q: Does the relation \eqref{TopologicalTorusObservables} represent the fundamental group of some parameter space $P$ associated with FQH systems?
\end{standout}

Remarkably --- the answer is \cite{SS25-FQH}: 
\begin{standout}
A: Yes, of the parameter space of cohomotopically quantized magnetic flux quanta on the torus.
\end{standout}

Namely, the parameter space of ordinary topological magnetic flux through a surface $\Sigma^2$, as predicted by the time-honored \emph{Dirac flux quantization condition} \cite{SS25-Flux}, is the space of maps
\begin{equation}
  \label{ConfigurationsOfMaxwellFields}
  P
  \defneq
  \mathrm{Map}\big(
    \Sigma^2
    ,
    B \mathrm{U}(1)
  \big)
  \,.
\end{equation}
to the classifying space 
\begin{equation}
  \label{MaxwellClassifyingSpace}
  \mathbb{C}P^\infty 
    \weakHomotopyEquivalence 
  B \mathrm{U}(1)
\end{equation}
for $\mathrm{U}(1)$ Maxwell gauge fields.
However this has abelian fundamental group \cite[Ex. 2.8]{SS25-FQH}
\begin{equation}
  \pi_1
  \,
    \mathrm{Map}\big(
    T^2
    ,
    B \mathrm{U}(1)
  \big)
  \simeq
  \mathbb{Z}^2
\end{equation}
whose representations are given by the algebras of Maxwell Wilson loop observables $W_a$, $W_b$, satisfying the undeformed commutation relation
\begin{equation}
  W_a W_b 
  =
  W_b W_a
  \,.
\end{equation}
On the other hand, \eqref{ConfigurationsOfMaxwellFields} describes, of course, only the free magnetic field and ignores its interaction with a 2D electron gas that we are concerned with here. Instead, the surplus magnetic flux quanta on a backdrop of a fractional filling fraction making an FQH system are associated with vortices in a strongly-coupled electron system and as such of exotic nature \cite[pp 883]{Stormer99}. To reflect this in their configuration topology we may ask for a deformation of the classical classifying space \eqref{MaxwellClassifyingSpace}. Since $\mathbb{C}P^\infty \defneq \bigcup_{n \in \mathbb{N}} \mathbb{C}P^n$, a suggestive choice of deformation is the first stage in this sequence (the ``3-skeleton''), which is the 2-sphere:
\begin{equation}
  \label{SphereCoefficientForFQH}
  \mathbb{C}P^1 \simeq S^2
  \mathrlap{\,.}
\end{equation}
This leads to an exotic topological flux parameter space being the \emph{Cohomotopy cocycle space} of $\Sigma^2$:
\begin{equation}
  \label{CohomotopyCocyclesAsParameters}
  P
  \defneq
  \mathrm{Map}\big(
    \Sigma^2
    ,
    S^2
  \big)
  \,.
\end{equation}
Remarkably, the fundamental group of this space in the connected component of winding/Chern number $C$ is (highlighted in \cite{SS25-FQAH}\cite[\S 3]{SS25-FQH}\cite[\S 3]{KSS26-HigherDimAnyons}, following \cite{Hansen1974,LarmoreThomas1980,Kallel2001}, more details below in \S\ref{TopOrderForNoSymmetry}) the nonabelian \emph{integer Heisenberg group} at level=2 (Def. \ref{IntegerHeisenbergGroup}):
\begin{equation}
  \pi_1
  \Big(
  \mathrm{Map}\big(
    \Sigma^2
    ,
    S^2
  \big),
  C
  \Big)
  \simeq
  H_3(\mathbb{Z}, 2C)
  \mathrlap{\,,}
\end{equation}
whose representations are exactly of the form \eqref{TopologicalTorusObservables}!

This shows in an otherwise familiar case how anyonic topological order can indeed arise from adiabatic monodromy in parameter spaces (here: \eqref{CohomotopyCocyclesAsParameters}) other than the usual configuration spaces of points from Fig. \ref{BraidRepresentationByTransport}.
\end{example}

We are thus ready to understand how to identify potential anyonic topological order in FQAH systems. The first step is to identify the space $P$ of classical external parameters of these systems:

\begin{example}[The situation for FQAH systems]

The parameters that define an FCI crystal are just its crystal properties: The \emph{bindings} if described by a tight-binding model, and generally the potential seen by single electrons in the background of the crystal lattice, as encoded in their position-space Hamiltonian. But, under Fourier transform, it is exactly this data that is encoded by the Bloch Hamiltonian \eqref{TheBlochHamiltonian}, hence by the (equivariant) classifying map \eqref{EquivariantClassifyingMap}. We conclude that the external parameter space for crystalline FQAH systems is just the equivariant mapping space \eqref{SymmetricFragileBandTopologySpace}:
\begin{equation}
  P
  \defneq
  \mathrm{Map}\big(
    \BrillouinTorus
    ,
    S^2
  \big)^{\!G}
  \mathrlap{.}
\end{equation}

Here it is important to note that FQAH systems, like all topologically ordered systems, involve strongly interacting electrons, whose \emph{joint} Hilbert space $\HilbertSpace_{{}_{\ClassifyingMap}}$ of gapped ground states arises from and depends on the crystal stucture, and hence on the given single-electron Bloch Hamiltonian $H$,
in a complicated manner that is not well understood (like most strongly-interacting field theory is not). But what matters for our analysis is only that it \emph{does depend} on $\ClassifyingMap \in \mathrm{Map}\big(\BrillouinTorus, S^2\big)^{\!G}$: By the quantum adiabatic theorem (Fig. \ref{AdiabaticQuantumTransport}) this implies that \emph{if} the FQAH system is topologically ordered then its gapped ground state Hilbert spaces $\HilbertSpace$ \emph{do} represent the fundamental group of $\mathrm{Map}\big(\BrillouinTorus, S^2\big)^{\!G}$, in the connected component corresponding to the given topological phase \eqref{Equivariant2Cohomotopy}.
\end{example}

We conclude at this point that the possible topological order in 2D 2-band $G$-crystalline FCIs, in any given charge sector 
$$
  C_G \in
  \pi_0 
  \,
  \mathrm{Map}\big(
    \BrillouinTorus, 
    S^2
  \big)^{\!G}
  ,
$$
ought to be characterized by representations
\begin{equation}
  \label{HilbertSpaceAsRep}
  \HilbertSpace
  \in
  \mathrm{Rep}\Big(
    \pi_1\,
  \mathrm{Map}_{C_G}
  \big(
    \BrillouinTorus, 
    S^2
  \big)^{\!G}
  \Big)
  \,.
\end{equation}

Remarkably, the coefficient space $S^2$ for fragile band topologies appearing here coincides with the classifying space \eqref{SphereCoefficientForFQH} postulated for FQH systems in Ex. \ref{TraditionalTopologicalOrderOverTheTorus}. This means that, when all crystalline symmetry is broken ($G  = 1$), we predict that potential topological order of 2-band 2D fractional Chern insulators is of just the traditional form \eqref{TopologicalTorusObservables}
for topological order over a torus, except that this torus is now the Brillouin torus of crystal momenta (highlighted in \cite[Thm. 1]{SS25-FQAH})!

The evident question of what happens to these anyon states over momentum space in the presence of non-trivial crystalline symmetry is hence answered by computing not just the connected components \eqref{Equivariant2Cohomotopy} but also their fundamental groups \eqref{HilbertSpaceAsRep}.

This is the second set of results on which we will report in \S\ref{TopologicalOrder}.

\subsection{Covariance under Diffeomorphisms}
\label{CovarianceUnderDiffeomorphisms}

It thus remains to understand properly the relevant parameter spaces $P$ (Def. \ref{GeneralTopologicalOrder}) of potentially topologically ordered quantum systems. We have discussed key examples above, but there remains one subtlety to take care of:

For parameters which are topological fields on a domain $\Sigma^d$, in that they are (equivariant) maps to a coefficient space $\hotype{A}$ (as in Exs. \ref{TraditionalTopologicalOrderOverTheTorus} and \ref{TraditionalTopologicalOrderOverTheTorus})
\begin{equation}
  P 
  \defneq
  \mathrm{Map}\big(
    \Sigma^d
    ,
    \hotype{A}
  \big)^{\!G}
  ,
\end{equation}
it stands to reason that parameter pairs $p,p' : \Sigma^d \xrightarrow{\;} \hotype{A}$ which differ only by the action of a(n equivariant) diffeomorphism $\phi : \Sigma^d \xrightarrow{\sim} \Sigma^d$, in that 
\begin{equation}
  p'
  =
  p \circ \phi
  \,,
\end{equation}
are perceived as equivalent by the topological ground states depending on them. Indeed, this is necessarily the case if the topological ground state spaces $\HilbertSpace$ behave like state spaces of topological quantum field theories (TQFT \cite{Witten1988, nLab:TQFT}), as is reasonably and generally expected to be the case, cf. \cite{Simon2023}.

In this situation, much as in \eqref{UnitaryTransformationOverPath}, the Hilbert spaces associated with parameter pairs related by a diffeomorphism $\phi$ are not strictly equal but are unitarily equivalent, via an associated unitary transformation $U_\phi$:
  \begin{equation}
    \label{UnitaryTransformationOverDiffeo}
    \begin{tikzcd}[
      row sep=-2pt,
      decoration=snake
    ]
      \HilbertSpace_p
      \ar[
        rr,
        "{ U_\phi }"
      ]
      &&
      \HilbertSpace_{p'}
      \\
      p
      \ar[
        rr,
        |->,
        shorten=8pt,
        "{ 
          (-) \circ \phi
        }"
      ]
      &&
      p'
      \mathrlap{\,.}
    \end{tikzcd}
\end{equation}
Therefore, the correct parameter space for generally covariant topological quantum states must include \emph{paths of diffeomorphism}. The standard mathematical construction implementing this is the \emph{homotopy quotient} $(\mbox{-})\sslash (\mbox{-})$
\cite[(A33)]{supp} (also known as the \emph{Borel construction}) of the mapping space by the (equivariant) diffeomorphism group:
\begin{equation}
  \label{CovariantizedParameterSpace}
  P 
  \defneq
  \mathrm{Map}\big(
    \Sigma^d
    ,
    \hotype{A}
  \big)^{\!G}
  \!\! \sslash
  \mathrm{Diff}\big(
    \Sigma^d
  \big)^{\!G}
  .
\end{equation}

\begin{example}
  To see this effect at work, it is instructive to consider the extreme case where the monodromy in the parameter space itself is trivial, say because the classifying space of the band topology is contractible, $\mathcal{A} \simeq \ast$, so that the diffeomorphism covariance in \eqref{CovariantizedParameterSpace} dominates the topological behaviour of the system: In this case, the fundamental group of the covariantized parameter space is just the equivariant modular group \eqref{EquivariantMCG}:
  $$
    \begin{aligned}
    \pi_1
    \Big(
    \mathrm{Maps}\big(
      T^2
      ,
      \ast
    \big)^G
    \!\sslash 
    \mathrm{Diff}\big(
      T^2
    \big)^G
    \Big)
    &
    \simeq
    \pi_1\Big(
      \mathrm{Diff}(T^2)^G
    \Big)
    \\
    &
    \simeq
    \pi_0\Big(
      \mathrm{Diff}\big(
        T^2
      \big)^G
    \Big)
    \\
    & \defneq
    \mathrm{Mod}^G
    \mathrlap{,}
    \end{aligned}
  $$
  and hence the possible anyonic topological order falls into modular representations. 

  In the general case the equivariant modular group is (\cite[Prop. 2.24]{SS25-FQH}) a semidirect factor of the covariantized monodromy:
  \begin{equation}
    \label{SemdirectProductMonodromy}
    \begin{aligned}
    &
    \pi_1
    \Big(
    \mathrm{Maps}\big(
      T^2
      ,
      \mathcal{A}
    \big)^G
    \!\sslash 
    \mathrm{Diff}\big(
      T^2
    \big)^G
    \Big)
    \\
    & \simeq
    \pi_1
    \Big(
    \mathrm{Maps}\big(
      T^2
      ,
      \mathcal{A}
    \big)^G
    \Big)
    \rtimes
    \mathrm{Mod}^G
    \mathrlap{.}
    \end{aligned}
  \end{equation}
\end{example}

In conclusion, the above analysis provides a substantial mathematical answer to the question raised in \S\ref{OnTopologicalOrderOfFCIs}, concerning the nature of possible anyonic topological quantum states in crystalline 2D fractional Chern-insulators: 

\begin{standout}
Topological order of crystalline 2-band FCIs falls into irreducible representations of the covariantized band monodromy \eqref{SemdirectProductMonodromy}, for $\mathcal{A} = S^2$ the classifying space for 2-Cohomotopy.
\end{standout}

We proceed now to working this out.

\section{Methods and Results}
\label{MethodsAndResults}

With the discussion in \S\ref{MotivationAndIntroduction}, we have reduced the physical question to the purely algebro-topological problem of computing the homotopy of the covariantized equivariant band topology spaces \eqref{CovariantizedParameterSpace}, hence computing:
\begin{description}
\item[\S\ref{TopologicalPhases}] 
their connected components $\pi_0$ \eqref{CovariantizedPhasesInIntro} (equivariant 2-Cohomotopy) for the possible topological phases, 
\item[\S\ref{TopologicalOrder}]
their fundamental groups $\pi_1$ \eqref{CovariantizedMonodromyInIntroduction} for the possible topological order.
\end{description}

Here we highlight interesting and illustrative examples of these computations, while a more comprehensive account is relegated to \cite{supp}.

\subsection{Generalities}

Our key move for dealing with the equivariant (crystalline) aspect is to determine minimal \emph{equivariant CW-complex structure} (cf. \cite[\S II]{supp}) on the Brillouin torus, for each given point group symmetry (examples are shown in Figs. \ref{StandardCellDecompositionOfPlainTorus} and \ref{p3CellStructure}) and then compute the homotopy of the equivariant mapping space by iterating over its skeleta  (cf. \cite[\S IV.B]{supp}). While as such this is a standard kind of move in equivariant homotopy theory, it seems not to have been applied to topological materials before, nor does the equivariant 2-Cohomotopy of 2-tori appear to have been computed before, by this or other means.

In contrast, for the $G$-action on the codomain 2-sphere (cf. \cite[\S III]{supp}) it turns out to only matter whether it fixes all points (case ``0'' of the trivial group action), two points (case ``I'' of cyclic group action) or no point (case ``II'' of dihedral group actions).

\begin{figure}[htb]
\caption{
\label{TheFixedPoles}
The action on the 2-sphere of the non-trivial cyclic groups $\CyclicGroup{n}$, $n \geq 1$, is by rotation around a fixed axis and hence fixes precisely an antipodal pair of points, to be denoted ``$\mathrm{n}$'' and ``$\mathrm{s}$''.
}
\centering
\adjustbox{
  rndfbox=4pt
}{
\begin{tikzpicture}
 \def\radius{1.4}

  \draw[
    dashed,
    line width=1
  ]
    (0,\radius*1.4) --
    (0,-\radius*1.5);

 \draw[
   white,
   line width=3,
   shift={(0,-\radius*1.2)}
 ]
   (-180-70:.5 and .15) arc
   (-180-70:-40: .5 and .15);
 \draw[
   -Stealth,
   shift={(0,-\radius*1.2)}
 ]
   (-180-70:.5 and .15) arc
   (-180-70:-40: .5 and .15);

  \shade[
    ball color=gray!40!white, 
    opacity=0.8
  ] (0,0) circle (\radius);

 \begin{scope}
   \clip (0,0) circle (\radius);
   \fill[
     blue,
     shift={(0,\radius*.997)}
   ]
     (0,0) ellipse 
       (.12 and .05);
   \fill[
     blue,
     shift={(0,-\radius*.997)}
   ]
     (0,0) ellipse
      (.12 and .05);
 \end{scope}

 \node[
   scale=.8
 ] at (85:\radius*1.07) {$\mathrm{n}$};
 \node[
   scale=.8
 ] at (-85:\radius*1.07) {$\mathrm{s}$};

\end{tikzpicture}
}

\end{figure}
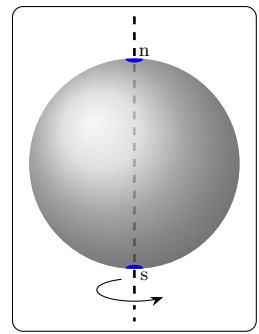

Here we focus on case (0 and) I, cf. Fig. \ref{TheFixedPoles}.

\subsection{Topological phases}
\label{TopologicalPhases}

Here we discuss instances of fragile topological phases of crystalline Chern insulators, namely of the connected components \eqref{CovariantizedPhasesInIntro}.

\subsubsection{No Symmetry}

\begin{figure}[htbp]
\caption{
  \label{StandardCellDecompositionOfPlainTorus}
  Minimal cell structure on the torus without symmetry, corresponding to the trivial 2D space group \SpaceGroup{p1}.
}
\centering
\adjustbox{
  rndfbox=4pt
}{
\begin{tabular}{c}
$
  \begin{tikzcd}[row sep=-3pt, column sep=0pt]
    \mathbb{Z}_{/1} \times (\mathbb{R}^2/\mathbb{Z}^2)
    \ar[
      rr,
      "{
        \SpaceGroup{p1}
      }"
    ]
    &&
    (\mathbb{R}^2/\mathbb{Z}^2)
    \\
    \big(
      [1]
      ,
      [x,y]
    \big)
    &\longmapsto&
    {[x,y]}
  \end{tikzcd}
$
\\
\\
\begin{tikzpicture}[
  scale=1.3,
  baseline=(current bounding box.center),
]

\fill[
  olive!30
]
 (-.8,-.8) rectangle
 (+.8,+.8);

\node[scale=2] at (0,0) 
  {$\circlearrowleft$};

\draw[
  line width=1.3,
  darkorange
]
  (-.8, -.8) --
  (-.8,+.8);
\draw[
  line width=.7,
  -Latex
]
  (-.8, .15) --
  (-.8, .15+.01);
\draw[
  line width=1.3,
  darkorange
]
  (+.8, -.8) --
  (+.8,+.8);
\draw[
  line width=.7,
  -Latex
]
  (+.8, .15) --
  (+.8, .15+.01);

\draw[
  line width=1.3,
  darkgreen
]
  (-.8, +.8) --
  (+.8,+.8);
\draw[
  line width=.7,
  -Latex
]
  (.1,+.8) --
  (.1+.01,+.8);

\draw[
  line width=1.3,
  darkgreen
]
  (-.8, -.8) --
  (+.8,-.8);
\draw[
  line width=.7,
  -Latex
]
  (+.1,-.8) --
  (.1+.01,-.8);

\fill 
  (-.8,-.8) circle (.06);
\fill 
  (+.8,-.8) circle (.06);
\fill 
  (-.8,+.8) circle (.06);
\fill 
  (+.8,+.8) circle (.06);

\draw[
  -Latex,
  gray,
  line width=.8
] 
  (-.8-.5, -.8-.3) --
  (.8+.5, -.8-.3);

\draw[
  -Latex,
  gray,
  line width=.8
] 
  (-.8-.3, -.8-.5) --
  (-.8-.3, .8+.5);

\draw[
  gray,
  line width=.8
] 
  (0,-.8-.3+.1) --
  (0,-.8-.3-.1);
\draw[
  gray,
  line width=.8
] 
  (-.8,-.8-.3+.1) --
  (-.8,-.8-.3-.1);
\draw[
  gray,
  line width=.8
] 
  (+.8,-.8-.3+.1) --
  (+.8,-.8-.3-.1);

\draw[
  gray,
  line width=.8
] 
  (-.8-.3+.1,0) --
  (-.8-.3-.1,0);
\draw[
  gray,
  line width=.8
] 
  (-.8-.3+.1, -.8) --
  (-.8-.3-.1, -.8);
\draw[
  gray,
  line width=.8
] 
  (-.8-.3+.1, +.8) --
  (-.8-.3-.1, +.8);

\node[
  scale=.7,
  color=gray
] at 
  (0,-.8-.55) {$0$};
\node[
  scale=.7,
  color=gray
] at 
  (-.8,-.8-.55) {$-\tfrac{1}{2}$};
\node[
  scale=.7,
  color=gray
] at 
  (+.8,-.8-.55) {$+\tfrac{1}{2}$};

\node[
  scale=.7,
  color=gray
] at 
  (-.8-.55, 0) {$0$};
\node[
  scale=.7,
  color=gray
] at 
  (-.8-.55-.08, -.8) {$-\tfrac{1}{2}$};
\node[
  scale=.7,
  color=gray
] at 
  (-.8-.55-.08, +.8) {$+\tfrac{1}{2}$};

\end{tikzpicture}
\end{tabular}
\;
\adjustbox{
  raise=4pt
}{
  \def\arraystretch{1.5}
  \begin{tabular}{|lc|}
    \hline
    \multicolumn{2}{|l|}{Cell decomposition}
    \\
    \hline
    \hline
    $D^0 \times \mathbb{Z}_{/1}/\mathbb{Z}_{/1}$
    &
    $\mathcolor{black}{\bullet}$
    \\[+2pt]
    $D^1 \times \mathbb{Z}_{/1}/\mathbb{Z}_{/1}$
    &
    \begin{tikzpicture}[
      baseline={([yshift=-2pt]current bounding box.center)}      
    ]
     \draw[
       line width=1.3,
       darkorange
     ]
       (0,0) --
       (0,.6);
    \draw[
      line width=.6,
      -Latex
    ]
      (0,.43) --
      (0,.43+.01);    
    \end{tikzpicture}
    \\[+5pt]
    $D^1 \times \mathbb{Z}_{/1}/\mathbb{Z}_{/1}$
    &
    \begin{tikzpicture}[
      baseline={([yshift=-2pt]current bounding box.center)},
      rotate=-90
    ]
     \draw[
       line width=1.3,
       darkgreen
     ]
       (0,0) --
       (0,.6);
    \draw[
      line width=.6,
      -Latex
    ]
      (0,.43) --
      (0,.43+.01);    
    \end{tikzpicture}
    \\[+2pt]
    $D^2 \times \mathbb{Z}_{/1}/\mathbb{Z}_{/1}$
    &
    \begin{tikzpicture}[
      baseline={([yshift=-2pt]current bounding box.center)}      
    ]
        \fill[olive!30]
          (-.3,-.3) rectangle
          (+.3,+.3);
        \node[scale=1.5] at (0,0) 
          {$\circlearrowleft$};
      \clip (0,0) rectangle (0,-.4);
    \end{tikzpicture}
    \\
    \hline
  \end{tabular}
}
}
\end{figure}
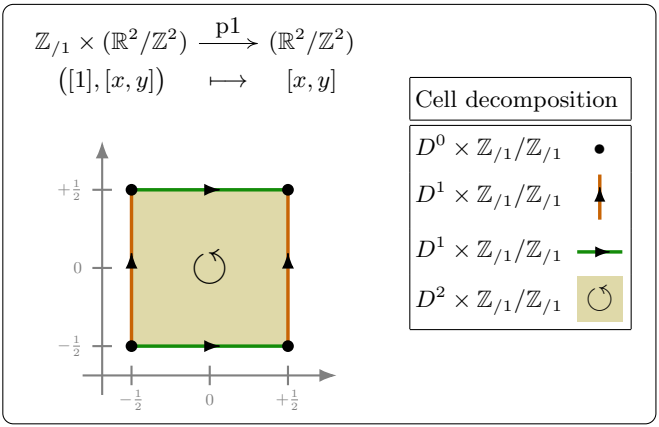

Just for completeness, we reiterate that in the case of trivial crystalline symmetry, the fragile topological phases in 2-Cohomotopy coincide with the stable Chern phases in ordinary integral 2-cohomology. One way to understand this is to observe that the inclusion of the respective classifying spaces
$$
  \begin{tikzcd}
    S^2 \simeq \mathbb{C}P^1
    \ar[r, hook, "{ i }"]
    &
    \mathbb{C}P^\infty
    \simeq
    B^2 \mathbb{Z}
  \end{tikzcd}
$$
is the projection onto the second Postnikov stage and hence induces an isomorphism on homotopy classes of maps out of any 2-manifold (cf. \cite[Prop. 1.20]{FSS23-Char}):

$$
  \begin{tikzcd}
    \pi^2(T^2)
    \ar[
      r,
      "{ \sim }"
    ]
    \ar[
      d,
      equals
    ]
    &
    H^2(T^2; \mathbb{Z})
    \ar[
      d,
      equals
    ]
    \\
    \pi_0, \mathrm{Map}(T^2, \mathbb{C}P^1)
    \ar[r, "{ i_\ast }"]
    &
    \pi_0, \mathrm{Map}(T^2, \mathbb{C}P^\infty)
    \mathrlap{\,.}
  \end{tikzcd}
$$
Moreover, on the left, the \emph{Hopf degree isomorphism}
$$
  \pi^2(T^2)
  \simeq
  \mathbb{Z}
$$
takes a map $\!\inlinetikzcd{T^2 \ar[r] \& S^2}\!$ to its \emph{winding number}, which may be thought of as the number of times that the 2-cell of the torus (shown in Fig. \ref{StandardCellDecompositionOfPlainTorus}) wraps around the 2-sphere. 

Since this winding number is manifestly invariant under precomposition with diffeomorphisms, it follows that in the absence of crystalline symmetry also the covariantized fragile classification \eqref{CovariantizedPhasesInIntro} is still by the usual integer (Chern) number:
\begin{equation}
  \pi_0
  \Big(
    \mathrm{Map}(T^2, S^2)
   \! \sslash 
    \mathrm{Diff}(T^2)
  \Big)
  \simeq
  \mathbb{Z}
  \mathrlap{\,.}
\end{equation}

But this default situation changes when the system is protected by crystalline symmetry, to which we turn next.

\subsubsection{\SpaceGroup{p3} Symmetry}
\label{OnTopologicalPhasesWithP3Symmetry}

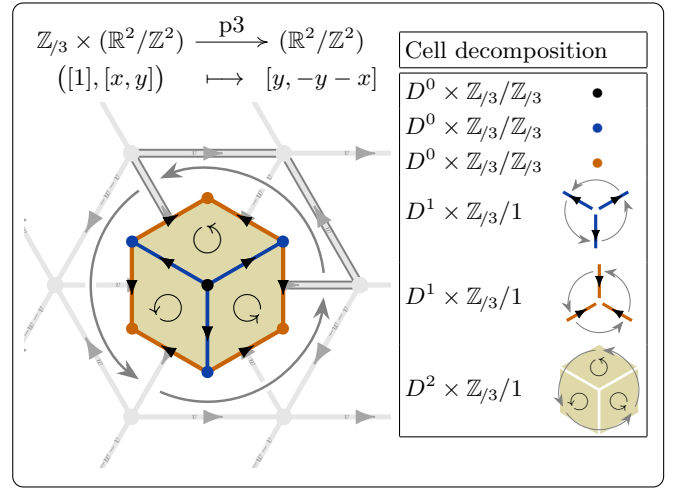
\begin{figure}[htb]
\caption{
  \label{p3CellStructure}
  Minimal equivariant cell structure on the torus with respect to the $\ZThree$-action corresponding to the space group \SpaceGroup{p3}.
}
\centering
\adjustbox{
  rndfbox=4pt
}{
\hspace{-11pt}
\begin{tabular}{c}
$
  \begin{tikzcd}[
    row sep=-3pt, 
    column sep=0pt]
    \ZThree \times
    (\mathbb{R}^2/\mathbb{Z}^2)    
    \ar[
      rr,
      "{ \SpaceGroup{p3} }"
    ]
    &&
    (\mathbb{R}^2/\mathbb{Z}^2)
    \\
    \big(
      [1]
      ,
      [x, y]
    \big)
    &\longmapsto&
    {[y, -y-x]}
  \end{tikzcd}
$
\\
\adjustbox{scale=1.35}{
\begin{tikzpicture}

\clip
  (-1.8,-1.8) rectangle
  (+1.8,+1.8);

\draw[
  gray,
  line width=2.5,
]
  (0,0) --
  (1.5,0) --
  ++ (120:1.5) --
  ++ (-1.5,0) --
  ++ (-60:1.5);

\TriVertex{0}{0}
\TriVertex{0}{1.5}
\TriVertex{60}{1.5}
\TriVertex{120}{1.5}
\TriVertex{180}{1.5}
\TriVertex{240}{1.5}
\TriVertex{300}{1.5}

\foreach \n in {0,1,2} {
\begin{scope}[rotate=\n*120]
\draw[
  line width=3,
  white
]
  (6:1.15) arc
  (6:120-6:1.15);
\draw[
  -Stealth,
  line width=.7,
  gray,
]
  (6:1.15) arc
  (6:120-6:1.15);
\end{scope}
}

\foreach 
  \n 
    [evaluate=\n as \angleone using {\n*120} ] 
    [evaluate=\n as \angletwo using {\n*120} ] 
  in {0,1,2} {

\begin{scope}[rotate=120*\n]

\fill[
  olive!30
]
  (0,0) --
  (30:.86) --
  (90:.86) --
  (150:.86) -- cycle;
\node[scale=1.3] at 
 (90:.45) {\rotatebox[origin=c]{\angleone}{$\circlearrowleft$}};

\draw[
  line width=1,
  color=darkblue
]
  (0,0) -- (30:.86);
\draw[
  line width=1,
  color=darkblue
]
  (0,0) -- (150:.86);
\draw[
  line width=.3,
  -Latex
]
  (30:.55) --
  (30:.55+.01);
\draw[
  line width=.3,
  -Latex
]
  (150:.55) --
  (150:.55+.01);

\draw[
  line width=1.2,
  color=darkorange
]
  (30:.86) -- (90:.86);
\draw[
  line width=.3,
  -Latex
]
  ($(90:.86)!.39!(30:.86)$)
  --
  ($(90:.86)!.38!(30:.86)$);

\draw[
  line width=1.2,
  color=darkorange
]
  (90:.86) -- 
  (150:.86);
\draw[
  line width=.3,
  -Latex
]
  ($(90:.86)!.39!(150:.86)$)
  --
  ($(90:.86)!.38!(150:.86)$);
  
\end{scope}
}

  \fill[black] 
    (0,0)
    circle (.06);
  \foreach \n in {0,1,2} {
  \begin{scope}[rotate=\n*120]
  \fill[darkorange] 
    ($(0:.33*1.5)+(120:.66*1.5)$)
    circle (.06);
  \end{scope}
}
\foreach \n in {0,1,2} {
  \begin{scope}[rotate=\n*120+60]
  \fill[darkblue] 
    ($(0:.33*1.5)+(120:.66*1.5)$)
    circle (.06);
  \end{scope}
}

\end{tikzpicture}
}
\end{tabular}
\hspace{-15pt}
\adjustbox{
  raise=4pt
}{
  \def\arraystretch{1.1}
  \begin{tabular}{|lc|}
    \hline
    \multicolumn{2}{|l|}{Cell decomposition}
    \\
    \hline
    \hline
    $D^0 \times \ZThree/\ZThree$
    &
    $\mathcolor{black}{\bullet}$
    \\
    $D^0 \times \ZThree/\ZThree$
    &
    $\mathcolor{darkblue}{\bullet}$
    \\
    $D^0 \times \ZThree/\ZThree$
    &
    $\mathcolor{darkorange}{\bullet}$
    \\
    $D^1 \times \ZThree/1$
    &
    \begin{tikzpicture}[
        baseline={([yshift=-2pt]current bounding box.center)}    
    ]
    \foreach \n in {0,1,2} {
      \begin{scope}[rotate=\n*120+30]
        \draw[
          line width=1.2,
          darkblue
        ]
          (0:.05) --
          (0:.5);
       \draw[
         -Latex,
         line width=.3
       ]
         (0:.4) --
         (0:.4+.01);
      \draw[
        -Stealth,
        gray
      ]
        (11:.41) arc
        (11:120-9:.41);
      \end{scope}
      }
    \end{tikzpicture}
    \\[+13pt]
    $D^1 \times \ZThree/1$
    &
    \begin{tikzpicture}[
        baseline={([yshift=-2pt]current bounding box.center)}    
    ]
    \foreach \n in {0,1,2} {
      \begin{scope}[rotate=\n*120+90]
        \draw[
          line width=1.2,
          darkorange
        ]
          (0:.05) --
          (0:.5);
       \draw[
         -Latex,
         line width=.3
       ]
         (0:.2) --
         (0:.2-.01);
      \draw[
        -Stealth,
        gray
      ]
        (11:.41) arc
        (11:120-9:.41);
      \end{scope}
      }
    \end{tikzpicture}
    \\[12pt]
    $D^2 \times \ZThree/1$
    &
    \begin{tikzpicture}[
        baseline={([yshift=-2pt]current bounding box.center)}        
    ]
    \fill[
       olive!30
    ]
      (-30:.6) --
      (30:.6) --
      (90:.6) --
      (150:.6) --
      (210:.6) --
      (270:.6) --
      (330:.6) -- cycle;
    \foreach 
      \n 
        [evaluate=\n as \angleone using {\n*120} ] 
        in {0,1,2} {
         \begin{scope}[rotate=\angleone]
            \node[scale=1.1] at 
             (90:.3) {\rotatebox[origin=c]{\angleone}{$\circlearrowleft$}};
         \end{scope}
         \draw[
           line width=1,
           white
         ]
          (0,0) --
          (\angleone+30:.6);
        }
    \foreach 
      \n in {0,1,2} {
            \draw[
              line width=.4,
              -Stealth,
              gray,
              rotate=\n*120
            ]
              (95:.53) arc
              (95:90+120-5:.53);
      }
    \end{tikzpicture}    
    \\
    \hline
 \end{tabular}
 }}
\end{figure}

Consider the 2D crystal symmetry \SpaceGroup{p3} (Fig. \ref{p3CellStructure}), where the dual crystal lattice $\mathbb{Z}^2 \subset \mathbb{R}^2$ is spanned by a pair of unit vectors enclosing an angle of $2\pi/3$, and where the point symmetry group $\ZThree$ is rotation by multiples of $2\pi/3$. Hence the group element $[1] \in \ZThree$ acts by taking one of the two basis vectors to the other and takes that latter basis vector to minus the sum of the two, whence its matrix representation in this basis is this:
\begin{equation}
  \label{MatrixRepresentationOfGeneratorOfP3PointSymmetry}
  L_{[1]}
  =
 \begin{pmatrix}
    0 & -1
    \\
    1 & -1
  \end{pmatrix}
  \,.
\end{equation}

The induced action of \eqref{MatrixRepresentationOfGeneratorOfP3PointSymmetry} on the torus $\BrillouinTorus = \mathbb{R}^2/\mathbb{Z}^2$ has exactly three fixed points $[p]$ with coordinate representatives
  \begin{equation}
    \label{RepresentativesOfTheP3FixedPoints}
    p_0
    :=
    \left(
    \begin{matrix}
      0
      \\
      0
    \end{matrix}
    \right)
    ,\;
    p_1
    :=
    \left(
    \begin{matrix}
      \tfrac{1}{3}
      \\
      \tfrac{2}{3}
    \end{matrix}
    \right)
    ,\;
    p_2 
    :=
    \left(
    \begin{matrix}
      \tfrac{2}{3}
      \\
      \tfrac{1}{3}
    \end{matrix}
    \right)
    .
  \end{equation}

Moreover, the typical symmetry transformation \eqref{TransformationOfHamiltonian}
of Bloch Hamiltonians on \SpaceGroup{p3} -Crystals 
involves $\ZThree$-rotation of the Bloch sphere, which has exactly two fixed points (cf. Fig. \ref{TheFixedPoles}).

Now, a $G$-equivariant map has to send $G$-fixed points (``high-symmetry points'') to $G$-fixed points. This immediately implies that the equivariant mapping space has $2^3 = 8$ sectors for the 8 ways of mapping the 3 high symmetry points to the 2 fixed poles on the sphere. But moreover, the $\ZThree$-equivariant diffeomorphisms of $T^2$ permute all three high symmetry points among each other (by \cite[Prop. B.1]{supp}), reducing the topologically distinguishable sectors to 
$$
  \left(\!\!{2 \choose 3}\!\!\right) 
  :=
  {{3 + 2 - 1} \choose 3} 
  =  
  4
  \mathrlap{\,.}
$$

In each of these sectors, the map from one of the three 2-cells in Fig. \ref{p3CellStructure} to $S^2$ has a winding number in $\mathbb{Z}$ (by \cite[\S IV B 6]{supp}), whence the total winding number is in $3 \mathbb{Z}$. Therefore the classes \eqref{CovariantizedPhasesInIntro} of covariantized fragile $\SpaceGroup{p3}$-crystalline phases are
\begin{equation}
  \label{p3TopologicalPhases}
  \pi_0\Big(
    \mathrm{Map}(T^2, S^2)^{\ZThree}
    \!\sslash
    \mathrm{Diff}(T^2)^{\ZThree}
  \Big)
  \simeq
  3 \mathbb{Z} \times [4]
  \mathrlap{\,.}
\end{equation}

In words: The fragile $\SpaceGroup{p3}$-crystalline phases are classified by pairs consisting of: 
\begin{enumerate}
\item[\bf (i)] a Chern class constrained to be divisible by 3,
\item[\bf (ii)] an element in the 4-element set.
\end{enumerate}
In particular, this predicts that the Chern class alone is no longer the unique classifier for $\SpaceGroup{p3}$-crystalline topological phases.

\subsection{Topological order}
\label{TopologicalOrder}

Here we discuss instances of potential topological order of 2D crystalline Chern insulators, namely of the fundamental groups \eqref{CovariantizedMonodromyInIntroduction} of their fragile band topology.

\subsubsection{No symmetry}
\label{TopOrderForNoSymmetry}

We sketch the proof (following \cite[\S 3]{SS25-Complete}\cite[\S 3.2]{KSS26-HigherDimAnyons}) of Ex. \ref{TraditionalTopologicalOrderOverTheTorus}, establishing the anyonic topological order 
for 2-cohomotopical charges
on the torus when all symmetry is broken.

To start with, recall that the plain torus admits a standard cell decomposition as shown in Fig. \ref{StandardCellDecompositionOfPlainTorus}.
Its 1-skeleton is the \emph{wedge sum} of two circles
\begin{equation}
  \label{1SkeletonOfPlainTorus}
  \mathrm{sk}_1(T^2)
  \simeq
  S^1 \vee S^1
  =
  \left\{
\begin{tikzpicture}[
  baseline={([yshift=-2pt]current bounding box.center)}
]

  \draw[
    line width=1.4,
    darkblue
  ]
  (0,0) -- (1.4,0);
  \draw[
    -Latex,
    line width=.7
  ]
    (.9,0) -- 
    (.9+.01,0);
  \draw[
    line width=1.4,
    darkblue
  ]
  (0,1.4) -- (1.4,1.4);
  \draw[
    -Latex,
    line width=.7
  ]
    (.9,1.4) -- 
    (.9+.01,1.4);

  \draw[
    line width=1.4,
    darkorange
  ]
  (0,0) -- (0,1.4);
  \draw[
    -Latex,
    line width=.7
  ]
    (0,.9) -- 
    (0,.9+.01);

  \draw[
    line width=1.4,
    darkorange
  ]
  (1.4,0) -- (1.4,1.4);
  \draw[
    -Latex,
    line width=.7
  ]
    (1.4,.9) -- 
    (1.4,.9+.01);

\fill
  (0,0) circle (.06);
\fill
  (1.4,0) circle (.06);
\fill
  (0,1.4) circle (.06);
\fill
  (1.4,1.4) circle (.06);
    
\end{tikzpicture}
  \right\}
\end{equation}
and the 2-cell attachment is the pushout
\begin{equation}
  \label{Standard2CellAttachmentForPlainTorus}
  \begin{tikzcd}[row sep=small]
    S^1
    \ar[
      rr,
      "{
        \circlearrowleft
      }"
    ]
    \ar[d]
    &
    {}
    \ar[
      dr,
      phantom,
      "{\ulcorner}"{pos=.9}
    ]
    &
    \mathrm{sk}_1(T^2)
    \ar[d]
    \\
    D^2 
    \ar[rr]
    &
    {}
    &
    T^2
    \mathrlap{\,,}
  \end{tikzcd}
\end{equation}
where the top map picks the boundary of the 2-cell as indicated in  Fig. \ref{StandardCellDecompositionOfPlainTorus}.

Now, each of the two 1-cells in \eqref{1SkeletonOfPlainTorus} contributes a copy of the integers to the fundamental group of the pointed mapping space, via 
$$
  \pi_1 \mathrm{Map}^\ast(S^1, S^2)
  \simeq
  \pi_0
  \mathrm{Map}^\ast(\Sigma S^1, S^2)
  \simeq
  \pi_2(S^2)
  \simeq
  \mathbb{Z}
$$
and so does the 2-cell \eqref{Standard2CellAttachmentForPlainTorus} now via the \emph{Hopf fibration} which generates $\pi_3(S^2)$:
$$
  \pi_1 \mathrm{Map}^\ast(S^2, S^2)
  \simeq
  \pi_0
  \mathrm{Map}^\ast(\Sigma S^2, S^2)
  \simeq
  \pi_3(S^2)
  \simeq
  \mathbb{Z}
  \,.
$$
By \emph{stable splitting} \cite[4I.1]{Hatcher2002}, these contributions combine multiplicatively to the underlying set of the pointed mapping space (in any connected component), which yields:
$$
  \pi_1
  \mathrm{Map}^\ast\big(
    T^2, S^2
  \big)
  \simeq_{\mathrm{Set}}
  \mathbb{Z}^2 \times \mathbb{Z}
  \mathrlap{\,.}
$$
Labelling the generating elements here as
\begin{equation}
  \label{GeneratorsOfHeisenberg}
  \left.
  \begin{aligned}
    W_a 
    & := 
    \big(
      (1,0),0
    \big)
    \\
    W_b 
    & := 
    \big(
      (0,1),0
    \big)
    \\
    \zeta
    & :=
    \big(
      (0,0),1
    \big)
  \end{aligned}
  \right\}
  \in
  \mathbb{Z}^2 \times \mathbb{Z}
  \mathrlap{\,,}
\end{equation}
then classical homotopy theory identifies the group commutator (cf. Fig. \ref{TorusGroupCommutator})
$$
  [W_a, W_b]
  \defneq
  W_a W_b W_a^{-1} W_b^{-1}
  \;\in\;
  \pi_1, \mathrm{Map}^\ast\big(
    T^2, S^2
  \big)
$$
with what is known as the \emph{Whitehead product} of the above homotopy groups (cf. \cite[p. 476]{Whitehead1978}), which turns out to take a pair of generators to \emph{twice} the class of the Hopf fibration (cf. \cite[p. 495]{Whitehead1978}):
\begin{equation}
  \label{TheWhiteheadProduct}
  \begin{tikzcd}[
    row sep=-1pt,
    column sep=20pt
  ]
    \pi_2(S^2)
    \times
    \pi_2(S^2)
    \ar[
      rr,
      "{ [-,-]_{\mathrm{Wh}} }"
    ]
    &&
    \pi_3(S^2)
    \\
    (1,1)
    \ar[
      rr,
      |->,
      shorten=12pt
    ]
    && 
    2
    \mathrlap{\,.}
  \end{tikzcd}
\end{equation}
Hence we find the characteristic anyon relation \eqref{TopologicalTorusObservables}  
$$
  [W_a, W_b]
  =
  \zeta^2
  \;\in\;
  \pi_1\, \mathrm{Map}^\ast(T^2, S^2)
  \mathrlap{\,,}
$$
where the crucial power of 2 on the right reflects the above Whitehead product \eqref{TheWhiteheadProduct}.

Moreover, already by degree-arguments one sees that this is the only nontrivial group commutator among the generators \eqref{GeneratorsOfHeisenberg}, 
$$
  [W_{a/b}, \zeta]
  =
  \mathrm{e}
  \;\in\;
  \pi_1\, \mathrm{Map}^\ast(T^2, S^2)
  \mathrlap{\,,}
$$
meaning that the \emph{anyon braiding phase} generator $\zeta$ in \eqref{GeneratorsOfHeisenberg} is indeed central. 

In summary, this shows that the fundamental group of the \emph{pointed} mapping space is the level=2 integer Heisenberg group of ``Planck constant'' $h = 0$:

\begin{definition}
  \label{IntegerHeisenbergGroup}
  For $h \in \mathbb{Z}$, the level=2 \emph{integer Heisenberg group} $\HeisenbergGroup{h}$ has underlying set $\mathbb{Z}^2 \times \mathbb{Z}_{/h}$ with group operation
  \begin{equation}
    \label{HeisenbergGroupOperation}
    \begin{tikzcd}[
      sep=-3pt,
      ampersand replacement=\&
    ]
      \begin{aligned}
      & (\HeisenbergGroup{h})
      \\
      \times
      &
      (\HeisenbergGroup{h})      
      \end{aligned}
      \ar[rr]
      \&\&
      \HeisenbergGroup{h}      
      \\
      \left(
        \begin{aligned}
        & (a,b,[n]),
        \\
        & (a',b',[n'])
        \end{aligned}
      \right)
      \&\longmapsto\&
      \left(        
        \begin{aligned}
        & a + a',
        \\
        & b+b',
        \end{aligned}
        ,
        \left[
          \begin{aligned}
            & n + n' + 
            \\
            & a b' - a' b
          \end{aligned}
        \right]
      \right)
      \mathrlap{\,.}
    \end{tikzcd}
  \end{equation}
\end{definition}

To conclude, one just needs to translate this result from the pointed to the unpointed mapping space. Using another classical result of Whitehead  \cite[(3.4)${}^\ast$]{Whitehead1946}, the Whitehead product \eqref{TheWhiteheadProduct} enters again in a different guise, and shows that in the connected component of the unpointed mapping space with Chern class $C$ the braid-phase extension group gets quotiented to $\mathbb{Z}_{/2C}$, whence the end result is:
\begin{equation}
  \label{HeisenbergAsFundamentalMappingSpaceGroup}
  \pi_1
  \Big(
  \mathrm{Map}\big(
    T^2
    ,
    S^2
  \big)
  ,C
  \Big)
  \simeq
  \HeisenbergGroup{2C}
  \mathrlap{\,,}
\end{equation}
as claimed.

Finally, it is clear from this derivation that the action of the plain modular group 
$$
  \mathrm{Mod}
  :=
  \pi_0\big(
    \mathrm{Diff}(T^2)
  \big)
  \simeq
  \mathrm{SL}(2,\mathbb{Z})
$$
on this fundamental mapping space group is via the defining action of $\mathrm{SL}(2,\mathbb{Z})$ on the $\mathbb{Z}^2$-factor in \eqref{HeisenbergAsFundamentalMappingSpaceGroup}. 

Moreover, careful analysis \cite[\S 3.4]{SS25-FQH} shows that in the finite-dimensional irreducible representations of the resulting \emph{covariantized band monodromy} group \eqref{CovariantizedMonodromyInIntroduction},
\begin{equation}
  \label{CovariantizedHeisenbergAsFundamentalMappingSpaceGroup}
  \begin{aligned}
  &
  \pi_1
  \Big(
  \mathrm{Map}\big(
    T^2
    ,
    S^2
  \big)
  \sslash
  \mathrm{Diff}(T^2)
  ,C
  \Big)
  \\
  &
  \simeq
  \big(
  \HeisenbergGroup{2C}
  \big)
  \rtimes
  \mathrm{SL}(2,\mathbb{Z})
  \mathrlap{\,,}
  \end{aligned}
\end{equation}
the anyon braid phase observable $\zeta$ \eqref{GeneratorsOfHeisenberg} becomes constrained to be a root of unity, as seen in fractional quantum Hall systems.

\subsubsection{\SpaceGroup{p3} Symmetry}
\label{OnTopologicalOrderForp3Symmetry}

It turns out (shown in \cite{supp}) that the cohomotopical anyonic topological order of \S\ref{TopOrderForNoSymmetry} disappears as soon as any non-trivial crystalline symmetry is respected. But \emph{another} kind of anyonic topological order appears for some crystalline $G$-symmetries, where the anyons are tied to the high symmetry points (the fixed points of the $G$-action):

This is in fact an immediate consequence of our observation \eqref{CovariantizedMonodromyInIntroduction} of covariantized topological order, combined with the result of \cite[Prop. B 1]{supp}, used already in \S\ref{OnTopologicalPhasesWithP3Symmetry}, that the $\SpaceGroup{p3}$-equivariant modular group \eqref{EquivariantMCG} contains the permutation group $\mathrm{Sym}(3)$ on the set $(T^2)^{\ZThree} \simeq [3]$ \eqref{RepresentativesOfTheP3FixedPoints} of high-symmetry points (cf. again Fig. \ref{p3CellStructure}) as a subgroup:
\begin{equation}
  \begin{tikzcd}
    &
    \mathrm{Mod}^{\ZThree}
    \ar[d]
    \\
    \mathrm{Sym}(3)
    \ar[r, "{ \sim }"]
    \ar[ur, hook]
    &
    \mathrm{Aut}\big(
      (T^2)^{\ZThree}
    \big)
    \mathrlap{\,.}
  \end{tikzcd}
\end{equation}

Namely, by \eqref{CovariantizedMonodromyInIntroduction} this means that $\mathrm{Sym}(3)$ is also a subgroup of the  covariantized band monodromy group of which the gapped ground states must be linear representations.

In fact, we can say more: In the fragile $\SpaceGroup{p3}$-crystalline topological phases \eqref{p3TopologicalPhases} of the form
$$
  (0, \mathrm{cnst}_{\mathrm{n}/\mathrm{s}})
  \;\in\;
  3\mathbb{Z} \times [4]
$$
(where all three high-symmetry points are jointly mapped to the same fixed pole of the coefficient 2-sphere, cf. Fig. \ref{TheFixedPoles}), the band monodromy group contains the subgroup of the \emph{framed symmetric group} $\mathbb{Z}^3 \rtimes \mathrm{Sym}(3)$ of total framing a multiple of 3. This is the result of \cite[Ex. 3.59]{SS25-FQH}: 
The ``framing'' witnesses a ``topological spin''-observable now associated with each high-symmetry point, as otherwise known to be carried by FQH anyons (cf. \cite[(2)]{SS25-FQH}), and $\mathrm{Sym}(3)$ itself appears as a quotient of the \emph{braid group} of would-be motions of the three high-symmetry points around each other. 

With  \eqref{CovariantizedMonodromyInIntroduction}, this means that the gapped ground states of such systems transform under the ``parastatistical'' form $\mathrm{Sym}(3)$ of the braid group $\!\!\!\inlinetikzcd{\mathrm{Br}(3) \ar[r, ->>] \& \mathrm{Sym}(3)}\!\!\!$ on three anyonic strands! (Cf. \cite[Rem. 3.60]{SS25-FQH}.)

\begin{figure}[htbp]
\caption{%
  \label{CyclicPermutation}%
  The monodromy in the covariantized space of fragile band topologies of $\SpaceGroup{p3}$-crystalline 2D 2-band systems reflects diffeomorphism equivariance which implements permutations of the three high-symmetry points. Among these, cyclic permutations manifest as (topologically protected) \emph{quantum rotation gates} (cf. Prop. \ref{Unitary2DIrrepOfSym3}).
}
\centering
\adjustbox{
  rndfbox=4pt
}{
\begin{tikzpicture}[scale=0.85]

  \draw[line width=1.5]
    (.5,0) .. controls
    (.5,1) and
    (0,1) ..
    (0,2);
  \draw[line width=1.5]
    (1,0) .. controls
    (1,1) and
    (.5,1) ..
    (.5,2);
    
  \draw[
    line width=6,
    white
  ]
    (0,0) .. controls
    (0,1) and
    (1,1) ..
    (1,2);
  \draw[line width=1.5]
    (0,0) .. controls
    (0,1) and
    (1,1) ..
    (1,2);

\begin{scope}
  \draw
    (115:.18 and .08) arc 
    (110:418:.18 and .08);
\end{scope}
\begin{scope}[shift={(.5,0)}]
  \draw
    (115:.18 and .08) arc 
    (110:418:.18 and .08);
\end{scope}
\begin{scope}[shift={(1,0)}]
  \draw
    (115:.18 and .08) arc 
    (110:418:.18 and .08);
\end{scope}
\begin{scope}[shift={(0,2)}]
  \draw[line width=1.7, white]
    (0:.18 and .08) arc 
    (0:360:.18 and .08);
  \draw
    (0:.18 and .08) arc 
    (0:360:.18 and .08);
\end{scope}
\begin{scope}[shift={(.5,2)}]
  \draw[line width=1.7, white]
    (0:.18 and .08) arc 
    (0:360:.18 and .08);
  \draw
    (0:.18 and .08) arc 
    (0:360:.18 and .08);
\end{scope}
\begin{scope}[shift={(1,2)}]
  \draw[line width=1.7, white]
    (0:.18 and .08) arc 
    (0:360:.18 and .08);
  \draw
    (0:.18 and .08) arc 
    (0:360:.18 and .08);
\end{scope}

\end{tikzpicture}
}
\end{figure}
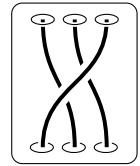

This is noteworthy:
\begin{proposition}[{\cite[Prop. 3.61]{SS25-FQH}}]
\label{Unitary2DIrrepOfSym3}
The unitary irrep $\mathbf{2}$ of $\mathrm{Sym}(3)$ is generated by a couple of \emph{quantum gates} \textup{(cf. \cite[\S 1.3]{NielsenChuang2010})} known as 
the \emph{Pauli Z-gate} $Z$,
and the \emph{rotation gate} $R_y(8\pi/3)$,
where the latter corresponds to the cyclic permutation of defect anyons \textup{(cf. Fig. \ref{CyclicPermutation})}.
\end{proposition}
Hence, if this \emph{topological order}, possible in $\SpaceGroup{p3}$-crystalline FQAH systems, could be experimentally controlled, it would implement topological protection for such quantum gates (cf. \cite[\S 3]{MySS2024}).

\section{Conclusion and Outlook}

We have first given a general theoretical re-assessment of the classification of fragile topological phases and of topological order specifically in crystalline 2-band systems, highlighting the previously possibly underappreciated facts that:
\begin{enumerate}
  \item[\bf (i)] 
  The fragile topological phases are classified by the \emph{equivariant 2-Cohomotopy} of the Brillouin torus, being the connected components of the covariantized space of band topologies,

  \item[\bf (ii)]
  the possible topological order is reflected in linear representations of the fundamental monodromy group of this covariantized space of band topologies.
\end{enumerate}
Neither of these have really been analyzed before. We have then used tools from equivariant homotopy theory --- which may not previously have found due attention in the solid state physics literature --- to systematically compute these effects. 

Finally, we have highlighted here noteworthy results of these computations for special choice of crystalline symmetry:
\begin{enumerate}
  \item[\bf (i)] The equivariant topological phases are generically labeled by pairs consisting of (a) a Chern number divisible by the order of the point group, and (b) an element of a finite set of ways that the high-symmetry points of the Brillouin torus map to fixed points in the coefficient 2-sphere.

  \item[\bf (ii)] When all symmetry is broken, then the adiabatic monodromy in the space of fragile band topologies reflects exactly the quantum observables of abelian FQH anyons on a torus --- here: localized not in position space but on the Brillouin torus of crystal momenta.

  \item[\bf (iii)] In the presence of crystalline symmetry these ``solitonic'' anyonic states disappear but ``defect'' anyonic states associated with the high-symmetry points appear, which, such as in the case of the $\SpaceGroup{p3}$-space group, exhibit non-abelian \emph{parastatistics}. These may serve to engineer topological protection for quantum gates such as the practically important rotation gates.
\end{enumerate}

All of these theoretical results appear to be new. As such,  they may help inform experimental searches for anyonic topological order in quantum materials, such as notably in FQAH systems, as recently highlighted in \cite{SS25-FQAH}.


\section*{Data Availability Statement}
This research is purely theoretical; therefore, no new data were created or analyzed in this study.

\begin{acknowledgments}
We thank the organizers of \emph{QMATH16} (Munich, Sept 2025), where the results reported here were first presented.

This research was supported by \emph{Tamkeen UAE} under the 
\emph{NYU Abu Dhabi Research Institute grant} {\tt CG008}.
\end{acknowledgments}

\bibliography{refs}


\end{document}